\documentclass[twocolumn,letterpaper,aps,prc,longbibliography,superscriptaddress,nofootinbib,floatfix]{revtex4-2}

\usepackage{graphicx}	
\usepackage{ltxtable}
\usepackage{amsmath}
\usepackage{verbatim}
\usepackage{xspace}	
\usepackage{enumitem}
\setlist[enumerate]{align=left}

\newcommand{\pt}{\mbox{$p_T$}\xspace}

\newcommand{\Ncoll}{\mbox{$N_{\mathrm coll}$}\xspace}

\newcommand{\ncoll}{\mbox{$N_{\mathrm coll}$}\xspace}
\newcommand{\meanncoll}{\mbox{$\langle N_{\mathrm coll} \rangle$}\xspace}

\newcommand{\psip}{\mbox{$\psi (2S)$}\xspace}

\newcommand{\sqsntwo}{\mbox{$\sqrt{s_{_{NN}}}=200~{\rm GeV}$}\xspace}

\newcommand{\sqstwo}{\mbox{$\sqrt{s}=200~{\rm GeV}$}\xspace}
\newcommand{\pp}{\mbox{$p$$+$$p$}\xspace}
\newcommand{\pA}{\mbox{$p$$+$$A$}\xspace}
\newcommand{\dau}{\mbox{$d$$+$Au}\xspace}

\newcommand{\pau}{\mbox{$p$$+$Au}\xspace}
\newcommand{\pal}{\mbox{$p$$+$Al}\xspace}
\newcommand{\ppb}{\mbox{$p$$+$Pb}\xspace}
\newcommand{\heau}{\mbox{$^{3}$He$+$Au}\xspace}

\newcommand{\jpsi}{\mbox{$J/\psi$}\xspace}

\newcommand{\pythia}{\mbox{\textsc{pythia8}}\xspace}

\begin{document}

\title{Measurement of $\psi(2S)$ nuclear modification 
at backward and forward rapidity in $p$$+$$p$, $p$$+$Al, and $p$$+$Au 
collisions at $\sqrt{s_{_{NN}}}=200$\,GeV}
 
 
\newcommand{\abilene}{Abilene Christian University, Abilene, Texas 79699, USA}
\newcommand{\augie}{Department of Physics, Augustana University, Sioux Falls, South Dakota 57197, USA}
\newcommand{\banaras}{Department of Physics, Banaras Hindu University, Varanasi 221005, India}
\newcommand{\barc}{Bhabha Atomic Research Centre, Bombay 400 085, India}
\newcommand{\baruch}{Baruch College, City University of New York, New York, New York, 10010 USA}
\newcommand{\bnlcoll}{Collider-Accelerator Department, Brookhaven National Laboratory, Upton, New York 11973-5000, USA}
\newcommand{\bnlphys}{Physics Department, Brookhaven National Laboratory, Upton, New York 11973-5000, USA}
\newcommand{\caucr}{University of California-Riverside, Riverside, California 92521, USA}
\newcommand{\charlesczech}{Charles University, Ovocn\'{y} trh 5, Praha 1, 116 36, Prague, Czech Republic}
\newcommand{\cns}{Center for Nuclear Study, Graduate School of Science, University of Tokyo, 7-3-1 Hongo, Bunkyo, Tokyo 113-0033, Japan}
\newcommand{\colorado}{University of Colorado, Boulder, Colorado 80309, USA}
\newcommand{\columbia}{Columbia University, New York, New York 10027 and Nevis Laboratories, Irvington, New York 10533, USA}
\newcommand{\czechtech}{Czech Technical University, Zikova 4, 166 36 Prague 6, Czech Republic}
\newcommand{\debrecen}{Debrecen University, H-4010 Debrecen, Egyetem t{\'e}r 1, Hungary}
\newcommand{\elte}{ELTE, E{\"o}tv{\"o}s Lor{\'a}nd University, H-1117 Budapest, P{\'a}zm{\'a}ny P.~s.~1/A, Hungary}
\newcommand{\eszterhazy}{Eszterh\'azy K\'aroly University, K\'aroly R\'obert Campus, H-3200 Gy\"ongy\"os, M\'atrai \'ut 36, Hungary}
\newcommand{\ewha}{Ewha Womans University, Seoul 120-750, Korea}
\newcommand{\famu}{Florida A\&M University, Tallahassee, FL 32307, USA}
\newcommand{\fsu}{Florida State University, Tallahassee, Florida 32306, USA}
\newcommand{\gsu}{Georgia State University, Atlanta, Georgia 30303, USA}
\newcommand{\hiroshima}{Hiroshima University, Kagamiyama, Higashi-Hiroshima 739-8526, Japan}
\newcommand{\howard}{Department of Physics and Astronomy, Howard University, Washington, DC 20059, USA}
\newcommand{\ihepprot}{IHEP Protvino, State Research Center of Russian Federation, Institute for High Energy Physics, Protvino, 142281, Russia}
\newcommand{\illuiuc}{University of Illinois at Urbana-Champaign, Urbana, Illinois 61801, USA}
\newcommand{\inrras}{Institute for Nuclear Research of the Russian Academy of Sciences, prospekt 60-letiya Oktyabrya 7a, Moscow 117312, Russia}
\newcommand{\instpasczech}{Institute of Physics, Academy of Sciences of the Czech Republic, Na Slovance 2, 182 21 Prague 8, Czech Republic}
\newcommand{\isu}{Iowa State University, Ames, Iowa 50011, USA}
\newcommand{\jaea}{Advanced Science Research Center, Japan Atomic Energy Agency, 2-4 Shirakata Shirane, Tokai-mura, Naka-gun, Ibaraki-ken 319-1195, Japan}
\newcommand{\jeonbuk}{Jeonbuk National University, Jeonju, 54896, Korea}
\newcommand{\kek}{KEK, High Energy Accelerator Research Organization, Tsukuba, Ibaraki 305-0801, Japan}
\newcommand{\korea}{Korea University, Seoul 02841, Korea}
\newcommand{\kurchatov}{National Research Center ``Kurchatov Institute", Moscow, 123098 Russia}
\newcommand{\kyoto}{Kyoto University, Kyoto 606-8502, Japan}
\newcommand{\lawllnl}{Lawrence Livermore National Laboratory, Livermore, California 94550, USA}
\newcommand{\losalamos}{Los Alamos National Laboratory, Los Alamos, New Mexico 87545, USA}
\newcommand{\lund}{Department of Physics, Lund University, Box 118, SE-221 00 Lund, Sweden}
\newcommand{\lyon}{IPNL, CNRS/IN2P3, Univ Lyon, Université Lyon 1, F-69622, Villeurbanne, France}
\newcommand{\maryland}{University of Maryland, College Park, Maryland 20742, USA}
\newcommand{\mass}{Department of Physics, University of Massachusetts, Amherst, Massachusetts 01003-9337, USA}
\newcommand{\michigan}{Department of Physics, University of Michigan, Ann Arbor, Michigan 48109-1040, USA}
\newcommand{\muhlenberg}{Muhlenberg College, Allentown, Pennsylvania 18104-5586, USA}
\newcommand{\nara}{Nara Women's University, Kita-uoya Nishi-machi Nara 630-8506, Japan}
\newcommand{\natmephi}{National Research Nuclear University, MEPhI, Moscow Engineering Physics Institute, Moscow, 115409, Russia}
\newcommand{\newmex}{University of New Mexico, Albuquerque, New Mexico 87131, USA}
\newcommand{\nmsu}{New Mexico State University, Las Cruces, New Mexico 88003, USA}
\newcommand{\northcg}{Physics and Astronomy Department, University of North Carolina at Greensboro, Greensboro, North Carolina 27412, USA}
\newcommand{\ohio}{Department of Physics and Astronomy, Ohio University, Athens, Ohio 45701, USA}
\newcommand{\ornl}{Oak Ridge National Laboratory, Oak Ridge, Tennessee 37831, USA}
\newcommand{\orsay}{IPN-Orsay, Univ.~Paris-Sud, CNRS/IN2P3, Universit\'e Paris-Saclay, BP1, F-91406, Orsay, France}
\newcommand{\peking}{Peking University, Beijing 100871, People's Republic of China}
\newcommand{\pnpi}{PNPI, Petersburg Nuclear Physics Institute, Gatchina, Leningrad region, 188300, Russia}
\newcommand{\pusan}{Pusan National University, Pusan 46241, Korea}
\newcommand{\riken}{RIKEN Nishina Center for Accelerator-Based Science, Wako, Saitama 351-0198, Japan}
\newcommand{\rikjrbrc}{RIKEN BNL Research Center, Brookhaven National Laboratory, Upton, New York 11973-5000, USA}
\newcommand{\rikkyo}{Physics Department, Rikkyo University, 3-34-1 Nishi-Ikebukuro, Toshima, Tokyo 171-8501, Japan}
\newcommand{\saispbstu}{Saint Petersburg State Polytechnic University, St.~Petersburg, 195251 Russia}
\newcommand{\seoulnat}{Department of Physics and Astronomy, Seoul National University, Seoul 151-742, Korea}
\newcommand{\stonybrkc}{Chemistry Department, Stony Brook University, SUNY, Stony Brook, New York 11794-3400, USA}
\newcommand{\stonycrkp}{Department of Physics and Astronomy, Stony Brook University, SUNY, Stony Brook, New York 11794-3800, USA}
\newcommand{\tenn}{University of Tennessee, Knoxville, Tennessee 37996, USA}
\newcommand{\texsu}{Texas Southern University, Houston, TX 77004, USA}
\newcommand{\titech}{Department of Physics, Tokyo Institute of Technology, Oh-okayama, Meguro, Tokyo 152-8551, Japan}
\newcommand{\tsukuba}{Tomonaga Center for the History of the Universe, University of Tsukuba, Tsukuba, Ibaraki 305, Japan}
\newcommand{\vandy}{Vanderbilt University, Nashville, Tennessee 37235, USA}
\newcommand{\weizmann}{Weizmann Institute, Rehovot 76100, Israel}
\newcommand{\wigner}{Institute for Particle and Nuclear Physics, Wigner Research Centre for Physics, Hungarian Academy of Sciences (Wigner RCP, RMKI) H-1525 Budapest 114, POBox 49, Budapest, Hungary}
\newcommand{\yonsei}{Yonsei University, IPAP, Seoul 120-749, Korea}
\newcommand{\zagreb}{Department of Physics, Faculty of Science, University of Zagreb, Bijeni\v{c}ka c.~32 HR-10002 Zagreb, Croatia}
\newcommand{\zambia}{Department of Physics, School of Natural Sciences, University of Zambia, Great East Road Campus, Box 32379, Lusaka, Zambia}
\affiliation{\abilene}
\affiliation{\augie}
\affiliation{\banaras}
\affiliation{\barc}
\affiliation{\baruch}
\affiliation{\bnlcoll}
\affiliation{\bnlphys}
\affiliation{\caucr}
\affiliation{\charlesczech}
\affiliation{\cns}
\affiliation{\colorado}
\affiliation{\columbia}
\affiliation{\czechtech}
\affiliation{\debrecen}
\affiliation{\elte}
\affiliation{\eszterhazy}
\affiliation{\ewha}
\affiliation{\famu}
\affiliation{\fsu}
\affiliation{\gsu}
\affiliation{\hiroshima}
\affiliation{\howard}
\affiliation{\ihepprot}
\affiliation{\illuiuc}
\affiliation{\inrras}
\affiliation{\instpasczech}
\affiliation{\isu}
\affiliation{\jaea}
\affiliation{\jeonbuk}
\affiliation{\kek}
\affiliation{\korea}
\affiliation{\kurchatov}
\affiliation{\kyoto}
\affiliation{\lawllnl}
\affiliation{\losalamos}
\affiliation{\lund}
\affiliation{\lyon}
\affiliation{\maryland}
\affiliation{\mass}
\affiliation{\michigan}
\affiliation{\muhlenberg}
\affiliation{\nara}
\affiliation{\natmephi}
\affiliation{\newmex}
\affiliation{\nmsu}
\affiliation{\northcg}
\affiliation{\ohio}
\affiliation{\ornl}
\affiliation{\orsay}
\affiliation{\peking}
\affiliation{\pnpi}
\affiliation{\pusan}
\affiliation{\riken}
\affiliation{\rikjrbrc}
\affiliation{\rikkyo}
\affiliation{\saispbstu}
\affiliation{\seoulnat}
\affiliation{\stonybrkc}
\affiliation{\stonycrkp}
\affiliation{\tenn}
\affiliation{\texsu}
\affiliation{\titech}
\affiliation{\tsukuba}
\affiliation{\vandy}
\affiliation{\weizmann}
\affiliation{\wigner}
\affiliation{\yonsei}
\affiliation{\zagreb}
\affiliation{\zambia}
\author{U.A.~Acharya} \affiliation{\gsu} 
\author{C.~Aidala} \affiliation{\michigan} 
\author{Y.~Akiba} \email[PHENIX Spokesperson: ]{akiba@rcf.rhic.bnl.gov} \affiliation{\riken} \affiliation{\rikjrbrc} 
\author{M.~Alfred} \affiliation{\howard} 
\author{V.~Andrieux} \affiliation{\michigan} 
\author{N.~Apadula} \affiliation{\isu} 
\author{H.~Asano} \affiliation{\kyoto} \affiliation{\riken} 
\author{B.~Azmoun} \affiliation{\bnlphys} 
\author{V.~Babintsev} \affiliation{\ihepprot} 
\author{N.S.~Bandara} \affiliation{\mass} 
% ...
\collaboration{PHENIX Collaboration}  \noaffiliation
 
\date{\today}

\begin{abstract}


Suppression of the $J/\psi$ nuclear-modification factor has been seen 
as a trademark signature of final-state effects in large collision 
systems for decades.  In small systems, the nuclear modification was 
attributed to cold-nuclear-matter effects until the observation of 
strong differential suppression of the $\psi(2S)$ state in 
$p/d$$+$$A$ collisions suggested the presence of final-state effects. 
Results of $J/\psi$ and $\psi(2S)$ measurements in the dimuon decay 
channel are presented here for $p$$+$$p$, $p$$+$Al, and $p$$+$Au 
collision systems at $\sqrt{s_{_{NN}}}=200$\,GeV.  The results are 
predominantly shown in the form of the nuclear-modification factor, 
$R_{pA}$, the ratio of the $\psi(2S)$ invariant yield per 
nucleon-nucleon collision in collisions of proton on target nucleus 
to that in $p$$+$$p$ collisions.  Measurements of the $J/\psi$ and 
$\psi(2S)$ nuclear-modification factor are compared with shadowing 
and transport-model predictions, as well as to complementary 
measurements at Large-Hadron-Collider energies.

\end{abstract}

\maketitle

 
\section{Introduction}
\label{sec:introduction}


Over the last decade, hydrodynamic calculations developed for $A$$+$$A$ 
collisions have been extended to small collision systems.  Recently, 
interest in small systems has surged with experimental data consistent 
with flow-like 
behavior~\cite{Nagle:2018nvi,CMS:2010ifv,ALICE:2012eyl,STAR:2016ydv,ATLAS:2017hap}, 
suggesting the possibility of hot-nuclear-matter effects in systems that 
were not previously believed to meet the threshold energy density for 
quark-gluon plasma formation~($\approx$1~GeV/fm$^3$).  In 2018, a PHENIX 
publication determined that elliptic and triangular-flow measurements in 
\pau, \dau, and \heau collisions of high event multiplicity were all 
consistent with hydrodynamic flow~\cite{PHENIX:2018lia}.  A more recent 
study performed using new analysis techniques has confirmed these 
results~\cite{PHENIX:2021bxz}.

In \dau collisions at the Relativistic Heavy Ion Collider (RHIC), 
preferential suppression of the quarkonia \psip state was observed, 
which is a possible signature of final-state 
effects~\cite{PHENIX:2013pmn}.  Suppression of the \psip nuclear 
modification factor was later observed at the Large Hadron Collider 
(LHC) by the ALICE and LHCb collaborations in \ppb 
collisions~\cite{ALICE:2014cgk,LHCb:2016vqr}.  The PHENIX \psip results 
were published following an analysis of \jpsi nuclear modification in 
\dau collisions, which indicated cold-nuclear-matter (CNM) effects were 
responsible for the modification of \jpsi 
production~\cite{PHENIX:2012czk}.  CNM effects are 
collectively known as any modification to charmonium production not 
caused by a hot and dense medium produced in the 
collision~\cite{Brambilla:2010cs}.  Modifications to the 
gluon-nuclear-parton distribution functions (nPDFs) in the 
nucleus~\cite{Eskola:2016oht,Kovarik:2015cma}, nuclear absorption 
(nuclear break up)~\cite{McGlinchey:2012bp,Arleo:1999af}, parton energy 
loss~\cite{Vitev:2007ve}, and the Cronin effect~\cite{Cronin:1974zm} are 
examples of CNM effects.  At LHC energies, similar \jpsi 
modification results were published in \ppb 
collisions~\cite{ALICE:2015kgk,ALICE:2013snh} and were also primarily 
consistent with CNM effects.

There is still an open debate within the community about the exact 
definition of initial- and final-state effects.  In particular, there is 
some ambiguity about whether nuclear absorption should be defined as an 
initial- or final-state effect. Throughout this paper, ``initial-state 
effects" are CNM effects, including nuclear absorption and ``final-state 
effects" are due to the energy produced during the collision.



The initial- and final-state effects are expected to be different at RHIC 
and LHC energies, so the comparison of PHENIX with LHC measurements is 
particularly valuable.  In general, \jpsi nuclear modification across the 
different experiments appears consistent with CNM effects, while the 
suppression observed in \psip nuclear modification is stronger with 
respect to the \jpsi nuclear modification than predicted by CNM effects.


Charmonium results have been published at LHC energies by the ATLAS, 
ALICE, LHCb, and CMS experiments in \ppb collisions at 
$\sqrt{s_{_{NN}}}=5.02$ TeV.  The ATLAS collaboration has reported 
\jpsi~\cite{Aad:2015ddl} and charmonium~\cite{Aaboud:2017cif} 
measurements.  The \jpsi~\cite{ALICE:2015kgk,ALICE:2013snh} and 
$\psi(2S)$~\cite{ALICE:2014cgk, ALICE:2016sdt} nuclear modification has 
been published by the ALICE collaboration, and more 
recently~\cite{Acharya:2020rvc} at $\sqrt{s_{_{NN}}}=8.16$ TeV.  The LHCb 
collaboration has measured \jpsi~\cite{LHCb:2013gmv} and 
$\psi(2S)$~\cite{LHCb:2016vqr}.  Lastly, \jpsi~\cite{Sirunyan:2017mzd} and 
$\psi(2S)$~\cite{Sirunyan:2018pse} measurements were published by the CMS 
collaboration.

At RHIC energies, \jpsi and \psip production was studied in \pp, 
\pal, \pau, \dau, and \heau collisions at \sqsntwo.  PHENIX has published 
\jpsi measurements at $|y|<0.35$ and $1.2<|y|<2.2$ in \pp 
collisions~\cite{PHENIX:2011gyb} and \dau 
collisions~\cite{Adare:2010fn,PHENIX:2012czk}.  At backward and forward 
rapidity $1.2 < |y| < 2.2$, PHENIX has published \jpsi nuclear 
modification in \pal, \pau, and \heau collisions~\cite{Acharya:2019zjt}.  
The \jpsi nuclear modification in \dau collisions was measured by the STAR 
collaboration at rapidity $|y|<1$~\cite{Adamczyk:2016dhc}.  The \psip 
nuclear-modification factor was also measured by PHENIX in \dau collisions 
at rapidity $|y| < 0.35$~\cite{PHENIX:2013pmn}, and the \psip to \jpsi 
ratio for the centrality-integrated case was measured in \pp, \pal, \pau, 
and \heau collisions at rapidity $1.2 < |y| < 2.2$~\cite{PHENIX:2016vmz}.  
The results presented in this paper provide the first measurements at RHIC 
of the \psip nuclear-modification factor and its centrality dependence at 
backward and forward rapidity.
 
 
\section{Experimental Setup}
\label{sec:experiment}

\subsection{The muon arms}

The PHENIX muon arms measure muons and unidentified-charged hadrons at 
backward and forward rapidity.  Covering the full azimuthal angle, the 
muon arms comprise four main components: The forward silicon vertex 
(FVTX), the muon tracker (MuTr), the muon identifier (MuID), and hadron 
absorbers~\cite{Akikawa:2003zs,Adachi:2013qha,Aidala:2013vna,Allen:2003zt}.

\begin{figure}[tbh]
\begin{center}
\includegraphics[width=1.0\linewidth]{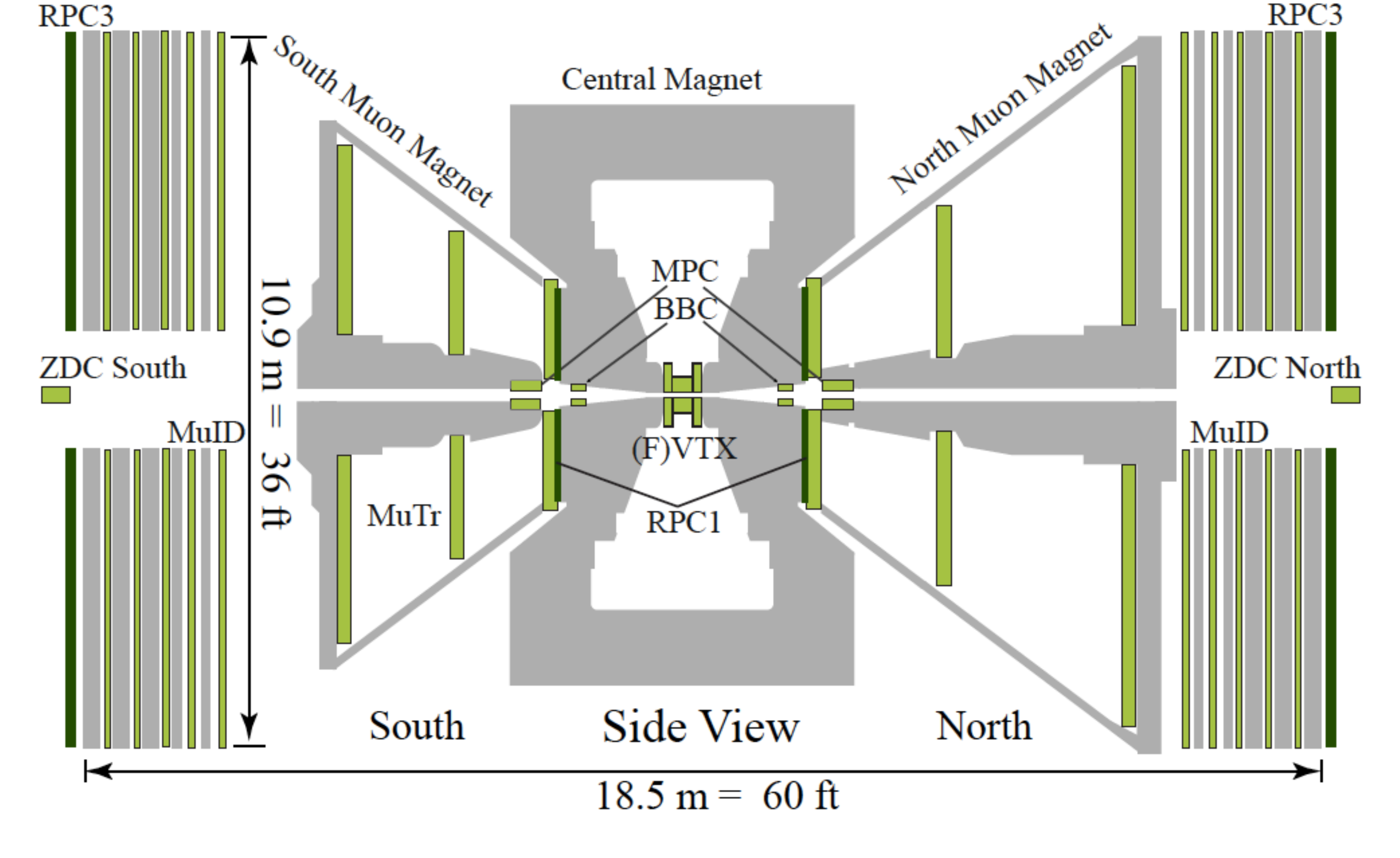}
\end{center}
\caption[Diagram of the PHENIX Detectors]{Side view of the PHENIX detector 
as configured for recording of data in 2015. }
\label{fig:phenix_sideview}
\end{figure}


Installed in 2012, the FVTX detector is a precision silicon detector 
comprising two identical endcaps containing four layers of active 
silicon sensors surrounding a barrel silicon vertex (VTX) detector.  As 
shown in Fig.~\ref{fig:phenix_sideview}, the FVTX is situated between 
the first hadron absorber and the collision region and provides an 
additional space point for muon-arms track reconstruction.  The location 
of the FTVX detector is critical, as particles traveling through the 
hadron absorbers experience multiple scattering, which impacts the mass 
resolution of dimuon pairs due to a less precise pair opening angle 
measurement.  The FVTX Detector used in the analysis is essential for 
extracting the \psip signal.


The PHENIX MuTr covers a pseudorapidity range of $-2.2<\eta<-1.2$ at 
backward rapidity and $1.2<\eta<2.4$ at forward rapidity. The MuTr 
comprises three individual cathode strip chambers called stations that 
face perpendicular to the direction of the beam.  Particles bend in the 
azimuthal direction under the $\oint{\textbf{B} \cdot d\textbf{l}}$ = 
0.72~T$\cdot$m magnetic field as they leave the interaction point.  
Signals induced on the cathode strips are used to reconstruct the 
trajectory of each particle.


The MuID is located outside the muon-arms magnetic field 
and directly behind the MuTr with respect to the collision vertex.  
Particles travel straight trajectories through five alternating layers of 
multi-wire chambers and hadronic absorbers called \textit{gaps}, where 
$gap~0$ is closest to the interaction point. The total hadronic absorber 
thickness along the beamline is 90~cm~(80~cm) in the MuID north~(south) 
arms.  A MuID absorber thickness of 90~cm corresponds to a 3\% probability 
for punch-through hadrons with maximum momenta of 4 GeV/$c$.


The hadron absorbers closest to the collision region are the 60~cm thick 
central magnet and 20~cm thick copper nose cones.  Two additional 
35-cm-thick stainless-steel absorbers run along the surface of the magnet 
at backward and forward rapidity. The MuID and the MuTr are divided by the 
muon-magnet yoke, which serves as a steel absorber 30~cm (20~cm) thick at 
forward (backward) rapidity.  The MuID includes an additional 80~cm 
of steel absorbers.  The total thickness of hadron-absorber material is 
225~cm (215~cm) in the north (south) muon arms.


Two beam-beam counters (BBC) are positioned inside an $\approx$0.3 T magnetic 
field region on both sides of the interaction point along the beamline.  
A \v{C}erenkov array detector, the BBC detectors comprise 128 identical 
quartz photo-multiplier tubes.  The acceptance for the BBC is full 
azimuthal angle and $3.1<|\eta|<3.9$ in pseudorapidity.  The BBC is used 
to determine the vertex position along the direction of the beam, and 
PHENIX also classifies the event centrality using the total charge 
recorded in the BBC.  In small systems, the centrality is determined using 
only the backward rapidity BBC charge.  The BBC was also used to measure 
the beam luminosity and form a minimum bias (MB) trigger.
Also shown in Fig.~\ref{fig:phenix_sideview} are the resistive-plate 
chambers (RPC) and the muon-piston calorimeter (MPC), used in 
geometry-related PHENIX measurements, which were not used in this 
analysis.



A Monte-Carlo-Glauber-model calculation is used to determine the 
centrality categorization. Values such as the inelastic nucleon-nucleon 
cross section and the nuclear charge density are input into the Glauber 
model, which simulates the probability of collision between nuclei based 
on the nucleon-nucleon inelastic cross section.  The total amount of 
charge in the BBC detector in the $A$-going direction ($-3.9<\eta<-3.1$) 
is used to define the centrality classes in small collision systems.  
\meanncoll, the mean number of binary (or nucleon-nucleon) collisions, for 
each centrality range are extracted from the Glauber 
Model~\cite{PHENIX:2013mpn}. The bias in the centrality measurement due to 
the presence of a hard process in the collision is accounted for by 
applying the $c_{\rm BBC}$ correction. Table~\ref{tab:centrality} lists the 
centrality classes and the \meanncoll and bias correction factors 
$c_{\rm BBC}$ for the \pal and \pau data sets.

\begin{table}[tbh]
\caption{
The mean number of
binary-binary collisions (\meanncoll) and bias correction factors
($c_{\rm BBC}$) determined from Glauber Model
calculations~\cite{PHENIX:2013mpn} for the different \pal and \pau 
collision-centrality classes.
}
\label{tab:centrality}
\begin{ruledtabular} \begin{tabular}{cccc}
 Collision system & Centrality & \meanncoll &  $c_{\rm BBC}$  \\
\hline
 \pal  & 0\%--100\% & 2.1$\pm$0.1  & 0.80$\pm$0.02 \\
\\
 \pau  & 0\%--20\%  & 8.2$\pm$0.5 & 0.90$\pm$0.01 \\
      & 20\%--40\% & 6.1$\pm$0.4 & 0.98$\pm$0.01 \\
      & 20\%--84\% & 4.3$\pm$0.3 & 1.00$\pm$0.03 \\
      & 40\%--84\% & 3.4$\pm$0.2  & 1.01$\pm$0.04 \\
      & 0\%--100\% & 4.7$\pm$0.3  & 0.86$\pm$0.01 \\  
\end{tabular} \end{ruledtabular}
\end{table}
 

\section{Data analysis}  
\subsection{Data Set}

Three different small system data sets were analyzed for the \psip 
measurements: \pp, \pal, and \pau collisions recorded during the 2015 run 
year, all at a center of mass energy of $\sqrt{s_{_{NN}}}$ = 200~GeV.  The 
corresponding integrated luminosities are 23~pb$^{-1}$, 260~nb$^{-1}$, and 
138~nb$^{-1}$ for the \pp, \pal, and \pau data sets, respectively.  
Forward rapidity results ($1.2 < y < 2.2$) correspond to the $p$-going 
direction, while backward rapidity results ($-2.2 < y < -1.2$) correspond 
to the Al/Au-going direction.  Analyzed \jpsi and \psip events were 
selected with a dimuon trigger that fires when two particles penetrate 
$gap~3$ in the MuID.
 
\subsection{Dimuon Track Selection}

The FVTX detector and the MuTr are both used for dimuon analysis 
in the PHENIX muon arms.  The \jpsi $\rightarrow \mu^+\mu^-$ decay channel 
has a large signal to background ratio, and the MuTr momentum resolution 
is sufficient for \jpsi analysis.  However, the MuTr momentum resolution 
is insufficient to extract the \psip yields, which are approximately 3\% 
of the \jpsi signal.  The FVTX detector provides additional space points 
near the collision vertex before the particle begins its trajectory 
through the muon-arm absorbers, enhancing the mass resolution.

A dimuon pair is composed of single muons (Muon1 and Muon2)
identified by the MuTr and FVTX detectors.  Four cases are listed below:

\begin{enumerate}[topsep=0pt,itemsep=-1ex,partopsep=1ex,parsep=1ex]

\item MuTr$+$MuTr (no FVTX--match requirement)

\item FVTX$+$MuTr (single FVTX--matched pair required)

\item FVTX$+$FVTX (double FVTX--matched pair required)

\item The sum of cases 2$+$3

\end{enumerate}

Case 1 was used in the recent \jpsi analysis with the same data sets for 
better statistics~\cite{Acharya:2019zjt}. Case 2, the single 
FVTX--matched pair, comprises a dimuon pair formed from one MuTr only 
track and one FVTX associated track.  Case 3, the double FVTX--matched 
pair, comprises a dimuon pair formed from two FVTX associated tracks.  
The \psip results presented in this paper have used Case 4, which is the 
sum of single+double FVTX-matched pairs.  

\subsection{\psip Signal Extraction}
\label{sec:data_signal}

The crystal-ball (CB) function~\cite{Gaiser:1982yw} and the modified-Hagedorn 
function~\cite{Aidala:2018ajl,Adare:2010de} were used to fit dimuon 
invariant mass distributions.  The CB function combines a 
power-law function with a Gaussian function~\cite{Gaiser:1982yw}.  The 
momentum resolution of the detector dominates the width of the \jpsi and 
\psip peaks, and the energy loss due to multiple scattering is modeled by 
the power-law tail.  The CB function along with the expanded 
parameters $A$ and $B$ for the power law tail is shown in 
Eq.~\ref{eq:cball}:

\begin{alignat}{3}
\label{eq:cball}
f(m) &= N\exp\left(-\frac{(m-\bar{m})^2}{2\sigma^2}\right) &&{\rm for}~ \frac{m-\bar{m}}{\sigma} > \alpha \\
f(m) &= NA\left(B-\frac{(m-\bar{m})^2}{2\sigma^2}\right) &&{\rm for}~ \frac{m-\bar{m}}{\sigma} \le \alpha \nonumber\\
A &= \left(\frac{n}{|\alpha|}\right)^n\exp\left(-\frac{|\alpha|^2}{2}\right), \nonumber \\
B &= \frac{n}{|\alpha|}-|\alpha|, \nonumber
\end{alignat}


The mass centroid and width of the \jpsi peak were allowed to vary in 
the centrality-integrated measurements.  To ensure fit stability in the 
finer centrality bins, the centroid and width were fixed to the 
centrality-integrated results for measurements as a function of 
\meanncoll, and the CB parameters $\alpha$ and $n$ were fixed to values 
determined from simulation (see Section~\ref{sec:signal_ext}).

\begin{figure*}[tbh]
\includegraphics[width=0.65\linewidth]{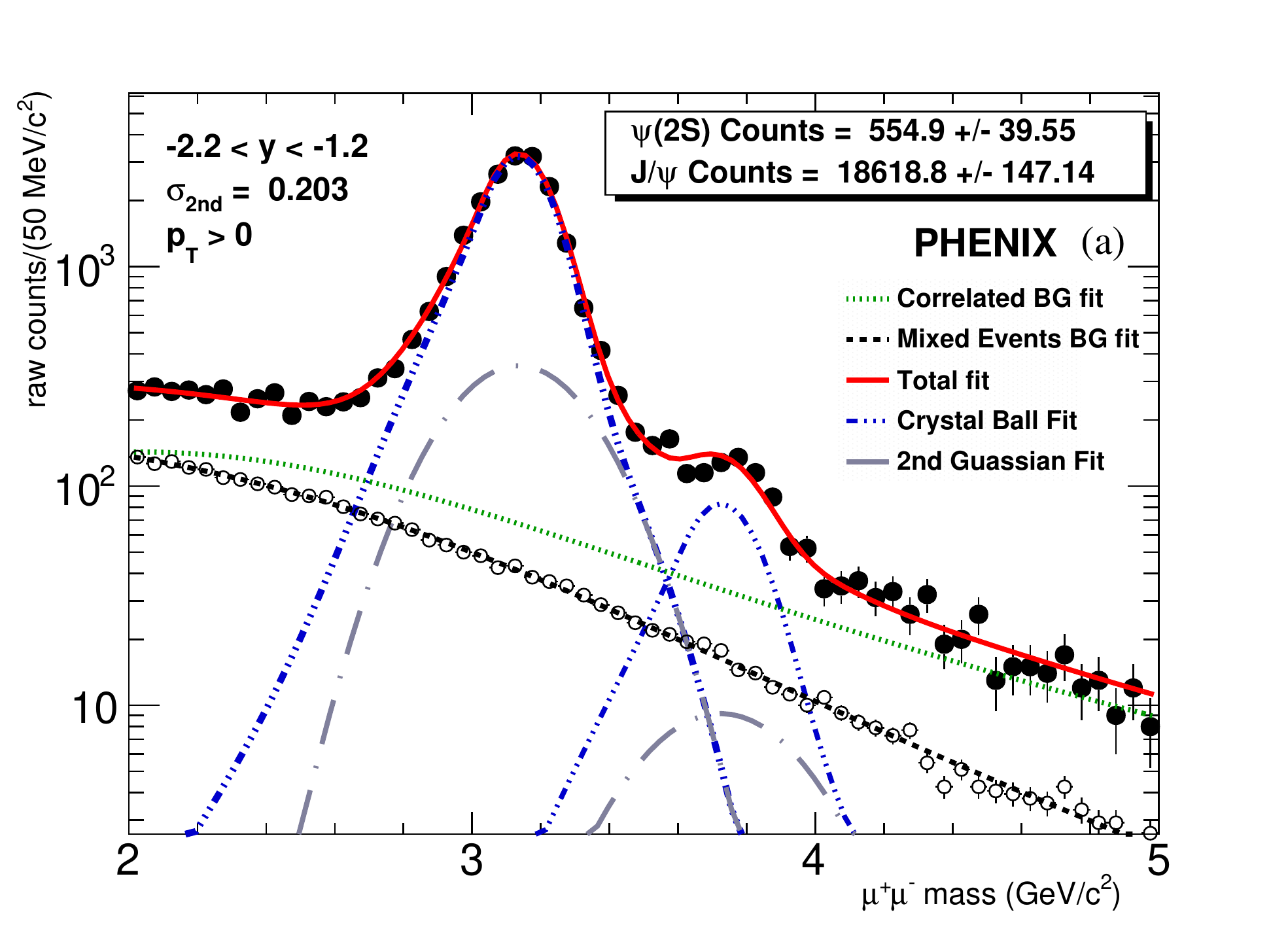}
\includegraphics[width=0.65\linewidth]{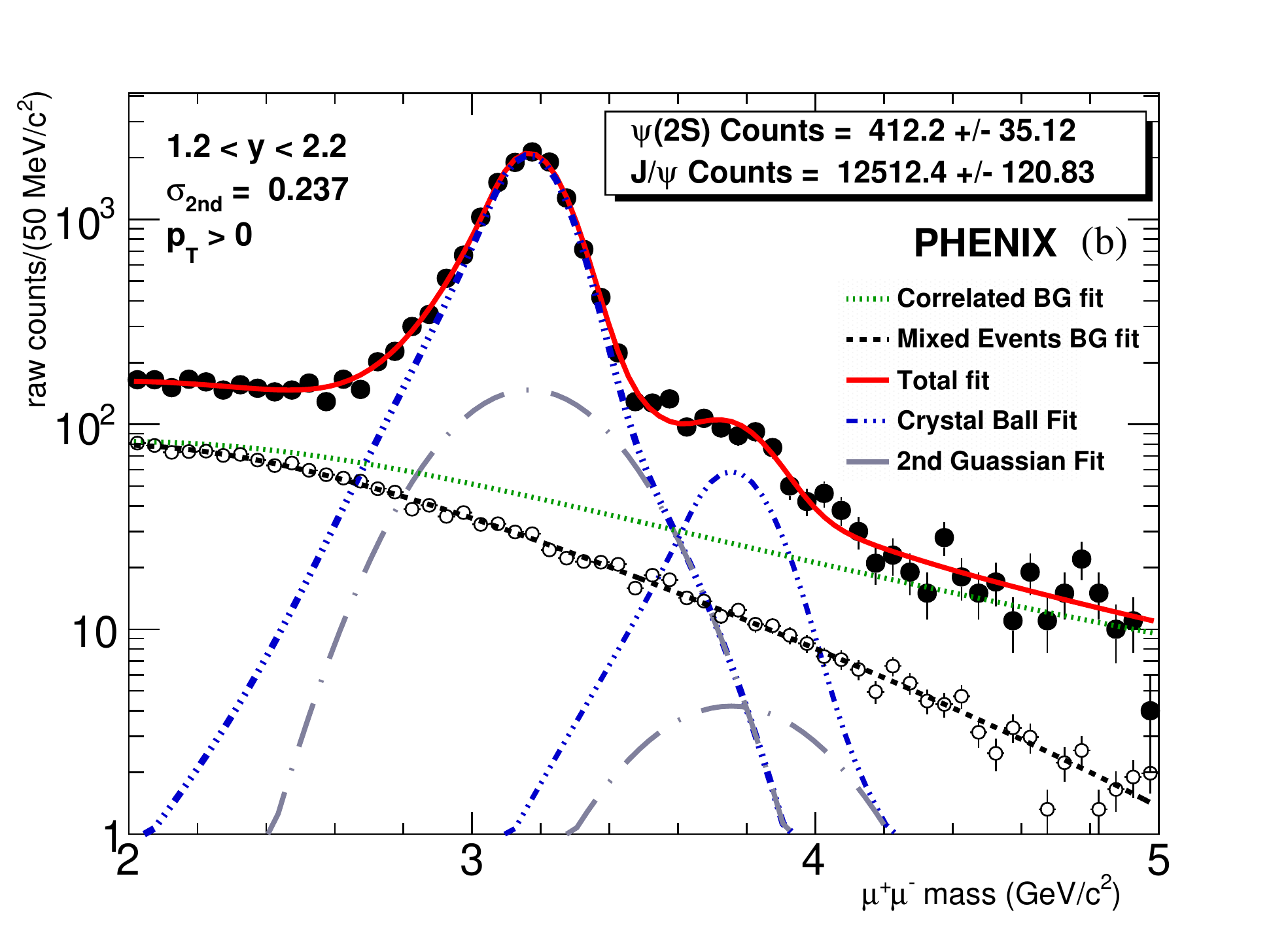}
\caption{\label{fig:ex_fit1} The \pt-integrated dimuon invariant mass 
spectrum in \pp collisions at (a) backward and (b) forward rapidity.  A 
CB function with a second Gaussian was used to extract the 
signals, and a modified-Hagedorn function was used to estimate the 
background contribution.}
\end{figure*}

Both the mass centroid and width of the \psip signal were fixed in all 
measurements.  The value of the \psip width was fixed to a result 
previously determined by simulation and based on the muon-arms mass 
resolution, where the ratio of the \psip to \jpsi width is expected to 
be 1.15~\cite{osti:912839}.  The \psip mass was fixed to the \jpsi mass 
plus a constant value $\Delta{m}$, as was done in a previous PHENIX 
analysis~\cite{PHENIX:2016vmz}.  The value $\Delta m$ is the mass 
difference between the \psip and \jpsi states, as reported by the 
Particle Data Group~\cite{PDG:2018}.  See Section~\ref{subsec:sys}
for a discussion of the systematic uncertainty associated with 
fixing the \psip lineshape.



In addition to the CB, a second Gaussian curve was used in the 
total fit function to reproduce the high mass tail seen in 
Fig.~\protect\ref{fig:ex_fit1}.  Misassociated tracks between the MuTr and 
the FVTX detectors can create this effect which is not observed in dimuon 
analyses where only MuTr information is used.  The parameters for the 
second Gaussian were determined in simulation and verified using a 
toy-model study that is discussed in more detail below.
 

The second Gaussian function contains three parameters: $\bar{m}_{\rm 2nd}$, 
$\sigma_{\rm 2nd}$, and $N_{\rm 2nd}$ which correspond to the mass centroid, 
width, and normalization of the curve. Based on the previous PHENIX 
muon-arms analysis~\cite{PHENIX:2016vmz}, the following second Gaussian 
parameter constraints are assumed:

  \begin{equation}
  \label{eq:constraints}
 f = f^\prime,  \hspace{5mm} r = r^\prime, \hspace{5mm} \bar{m} = \bar{m}_{\rm 2nd}, \hspace{5mm} \bar{m}^\prime = \bar{m}^\prime_{\rm 2nd},
 \end{equation}
where $f=\frac{N_{\rm 2nd}}{N}$ is the ratio of the \jpsi second Gaussian 
normalization to the CB normalization, $f^\prime$ = 
$\frac{N^\prime_{\rm 2nd}}{N^\prime}$ is the ratio of the \psip second 
Gaussian normalization to the CB normalization, 
$r = \frac{\sigma_{\rm 2nd}}{\sigma}$ is the ratio of the \jpsi second 
Gaussian width to the CB width, $r^{\prime} = 
\frac{\sigma^{\prime}_{\rm 2nd}}{\sigma^{\prime}}$ is the ratio of the \psip 
second Gaussian width to the CB width, $\bar{m}_{\rm 2nd}$ is the 
\jpsi second Gaussian mass centroid, $\bar{m}^\prime_{\rm 2nd}$ is the \psip 
second Gaussian mass centroid, $\bar{m}$ is the \jpsi CB mass 
centroid, and $\bar{m}^\prime$ is the \psip CB mass centroid.  
The ratios $f$ and $r$ were determined from fits to the simulated \jpsi 
mass distribution using embedded simulations with high statistics.  The 
ratios $f$ and $r$ were then used in real data analysis by multiplying 
them with the corresponding free parameter in the total fit function 
following the constraints given in Eq.~\ref{eq:constraints}.

To summarize the total fit function, there are fixed parameters included 
to stabilize the fit:  three parameters in the correlated background;  
two parameters (mean and sigma) in the \psip CB function; the 
ratio of mean and sigma between the CB function and the 2nd 
Gaussian function; $\alpha$ and $n$ in the CB tail parameters 
only for centrality dependent fits based on the parameters from the 
integrated-centrality fits. All fixed-parameter values are different for 
collision systems and rapidity ranges, except for the \psip mass and width 
parameter values.
 
\subsection{Background Estimation}
\label{sec:background}

A mixed-events background was generated to approximate the 
random-combinatorial-background contribution to the dimuon-invariant 
mass spectrum by selecting oppositely-charged single muons $\mu^+ \mu^-$ 
from different events.  An event pool of four events was used to reduce 
the statistical uncertainty of the mixed-events background.  All 
generated mixed events must be within 2~cm of the z-vertex and come from 
the same centrality class as the main event.

The correlated background comprises contributions from Drell-Yan, charm, 
bottom, and charged hadron dimuon pairs.  The correlated-background 
shape is not precisely known at the muon-arm acceptance.  The same 
approach was followed as in the recent analysis~\cite{Acharya:2019zjt}. 
A modified-Hagedorn function was included in the total fit function to 
estimate the correlated-background contribution, and the shape was 
constrained based on detailed simulation studies of all 
components~\cite{Aidala:2018ajl,Leung:2019vbb}. The modified Hagedorn is 
given as:

\begin{equation}
\label{eq:hag}
   \frac{d^2N}{dm_{\mu\mu}d\pt} = \frac{p_{0}}{[\exp{(-p_{1}m_{\mu\mu} - p_{2}m_{\mu\mu}^2)} + {m_{\mu\mu}}/{p_{3}}]^{p_{4}}}, 
\end{equation}
where $p_0$ is the normalization parameter, $p_4$ is the high mass tail 
parameter, $m_{\mu \mu}$ is the reconstructed dimuon mass, and $p_1$, 
$p_2$, and $p_3$ are fit parameters.  In the fits to the real data, the 
parameters $p_0$, $p_1$, and $p_3$ were fixed to values determined by 
the simulation studies, while the remaining parameters $p_2$ and $p_4$ 
were not fixed.  The systematic uncertainty associated with this 
approach is discussed in Section~\ref{subsec:sys}.

\subsection{Efficiency Correction}

The combined acceptance and reconstruction efficiency correction is 
applied to the dimuon invariant yield to compensate for the geometric 
acceptance of the detector and the track-reconstruction efficiency.  
This correction was determined using simulations, where \jpsi or \psip 
candidates were generated with \pythia~\cite{PYTHIA:81} and thrown into 
the geometric acceptance region of the PHENIX muon arms.  The simulated 
events were embedded into real physics data to account for the effects 
of background hits for $p$$+$$A$ collisions.  A full detector simulation 
was performed using \textsc{Geant4}~\cite{Geant:4}, and includes a 
dimuon trigger emulator for the trigger-efficiency determination.  The 
acceptance and reconstruction efficiency corrections for \psip in \pp 
collisions are $\approx$3.5\% and $\approx$4.5\% at backward and forward 
rapidity, respectively, for the requirement of at least one FVTX+MuTr 
matching.  In comparison, the \jpsi acceptance and reconstruction 
efficiency corrections in \pp collisions are $\approx$3.0\% and 
$\approx$3.5\% at the same rapidities.  The higher acceptance and 
reconstruction efficiency for the \psip is expected due to its larger 
mass.  In centrality-integrated \pau collisions, the \psip acceptance 
and reconstruction efficiency is $\approx$3.0\% and $\approx$4.0\% at 
backward and forward rapidities.

\subsection{Nuclear Modification Factor}

The dimuon invariant yield for a given rapidity, centrality, and \pt in a 
certain collision system is given as:

\begin{equation}
\label{eq:inv_yield}
    B_{\mu \mu} \frac{ d^2Y^{\mu \mu}}{dp_T dy} =  \frac{1} {2\pi \pt \Delta p_T \Delta y} \frac{ \epsilon_{\rm BBC} N^{\mu \mu}}{ \epsilon_{\rm trig} \epsilon_{\rm Ae} N_{\rm MB}},
\vspace{3mm} 
\end{equation} 
where $N^{\mu \mu}$ is the number of raw (uncorrected) \jpsi or \psip 
counts per bin, $B_{\mu \mu}$ is the branching ratio to dimuons, 
$\epsilon_{\rm trig}$ and $\epsilon_{\rm Ae}$ are the dimuon trigger and 
combined acceptance and reconstruction efficiencies, $p_T$ is the center 
of the $p_T$ bin, $\Delta p_T$ is the \pt bin width, $\Delta y$ is the 
rapidity bin width, and $N_{\rm MB}$ is the raw number of MB events.  
The BBC efficiencies for MB events and hard-scattering events are 
included in $\epsilon_{\rm BBC}$, and the list of correction factors for 
various centrality ranges in \pA collisions is provided in 
Table~\ref{tab:centrality}. For the \pp data set, the BBC efficiency for 
MB (hard-scattering) events is $0.55\pm0.05$ 
($0.79\pm0.02$)~\cite{Drees:2003zza}.

The nuclear-modification factor $R_{pA}$, the ratio of the dimuon 
invariant yields in \pA collisions compared to the dimuon yield in \pp 
collisions, scaled by the average number of nucleon-nucleon collisions, is 
the primary observable used in the analysis:

\begin{equation}
\label{eq:rab}
    R_{pA} = \frac{1}{\langle \ncoll 
\rangle}\frac{d^{2}N^{pA}/dydp_T}{d^{2}N^{pp}/dydp_T},
\end{equation}
where $\langle \ncoll \rangle$ is the average number of nucleon-nucleon 
collisions per event, and $d^2N^{pA}/dyd\pt$ and $d^2N^{pp}/dyd\pt$ are 
the dimuon invariant yields in the \pA and \pp systems, respectively.

\subsection{Systematic Uncertainties}
\label{subsec:sys}

The systematic uncertainties are divided into three different types:  
Type A, Type B, and Type C uncertainties.  Type A uncertainties are 
uncorrelated, random point-to-point uncertainties.  In this paper, the 
Type A uncertainties are associated with the extracted dimuon yields and 
are added in quadrature to statistical uncertainties.  Type B 
uncertainties are correlated point-to-point systematic uncertainties, and 
Type C uncertainties are global uncertainties that apply uniformly to all 
measurements.

\subsubsection{Signal Extraction}
\label{sec:signal_ext}

\begin{table}[tbh]
\caption{\label{tab:sys_fit1}
The \psip fractional systematic uncertainties for signal extraction in
\pp, \pal, and \pau collisions. ``CB" denotes the crystal-ball fit
function~\protect\cite{Gaiser:1982yw}.
}
\begin{ruledtabular} \begin{tabular}{ccccc}
 Source         & System  & Forward   & Backward  & Type \\ 
\hline
 Prob. Cut       & \pp    & 9.3\%    & 9.3\%       & B \\ 
\\
 Corr. Bkg.      & \pp    & 3.3\%    & 3.2\%    & B \\ 
                 & \pal   & 7.5\%    & 5.7\%    & B \\
                 & \pau   & 6.9\%    & 8.4\%    & B \\  
\\
 Comb. Bkg.      & \pp    & $<$1.0\% & $<$1.0\% & B \\ 
                 & \pal   & 1.0\%    & 1.0\%    & B \\
                 & \pau   & 1.00\%    & 3.7\%    & B \\  
\\
 Fixed CB Shape  & \pp    & 1.1\%    & 1.4\%     & B \\ 
                 & \pal   & 1.0\%    & 1.1\%    & B \\
                 & \pau   & 1.4\%    & 1.4\%    & B \\ 
\\
 Fixed CB Tail   & \pp    &    -      &    -      & B \\ 
                 & \pal   & -         & -         & B \\
                 & \pau   & 2.6\%    & 3.8\%    & B \\  
\\
 Fit Procedure   & \pp    & 3.3\%    & 3.3\%    & B \\ 
                 & \pal   & 2.0\%    & 2.7\%    & B \\
                 & \pau   & 4.7\%    & 3.3\%    & B 
\end{tabular} \end{ruledtabular}
\end{table}

\begin{table}[tbh]
\caption{\label{tab:sys_fit2}
The 0\%--100\% centrality \jpsi fractional systematic uncertainties for
the signal extraction in \pp, \pal, and \pau collisions.  The \jpsi
lineshape and the CB tail parameters were not fixed for
measurements in MB-triggered data.
}
\begin{ruledtabular} \begin{tabular}{ccccc}
Source & System       & Forward        & Backward        & Type\\
\hline
 Corr. Bkg.    & \pp    & $<$1.0\%  & $<$1.0\%     & B      \\ 
               & \pal   & $<$1.0\%  &   $<$1.0\%   & B       \\
               & \pau   & $<$1.0\%  & $<$1.0\%     & B   \\  
\\
 Comb. Bkg.    & \pp    & $<$1.0\%  & $<$1.0\%     & B      \\ 
               & \pal   & $<$1.0\%  & $<$1.0\%     & B     \\
               & \pau   & $<$1.0\%  & $<$1,0\%     & B   \\ 
\\
 Fit Procedure & \pp    &  1.2\%    &   1.0\%      & B    \\ 
               & \pal   & 1.0\%     & 1.3\%        & B        \\
               & \pau   & 1.1\%     & 1.4\%        & B    
\end{tabular} \end{ruledtabular}
\end{table}


After analyzing the \psip invariant yield results from three different 
sets of analysis cuts, an outlier measurement was observed in the \pp 
collision system.  The three different sets were defined using the 
following criteria: Set 1 has standard cuts for quality of track 
reconstruction in the MuTr, MuID, and FVTX applied, Set 2 has an 
additional fiducial cut applied, and Set 3 has an additional quality cut 
applied to the FVTX tracks.  Note that Data Set 3 was used for all 
results presented in this paper.  The outlier did not contain a 
probability cut, which reduces the number of misassociated tracks 
between the FVTX and MuTr detectors. The probability cut 
\cite{PHENIX:2017caf} is an FVTX track $\chi^{2}$ probability (p-value) 
quality selection cut; a probability cut of greater than 3\% was applied 
to data Set 3.  To account for the observed discrepancy in invariant 
yield results, the weighted averages of invariant yields from each muon 
arm in the three sets were compared, and the largest percent difference 
between them was taken as a systematic uncertainty.  No outliers were 
observed in the comparison of the three measurements in the \pal or \pau 
collision systems, and no systematic uncertainty was assigned.


Aside from the probability cut, the dominant uncertainty in the \psip 
measurements is the correlated-background uncertainty.  The modified 
Hagedorn function shown in Eq.~\ref{eq:hag} was used to fit the 
estimated correlated background, as was done in a recent PHENIX 
analysis~\cite{Acharya:2019zjt}.  The shape was determined from 
simulation studies~\cite{Aidala:2018ajl, Leung:2019vbb}, and two of the 
five parameters in the fit function were allowed to vary.  The 
systematic uncertainty was determined from three separate Hagedorn fits, 
with parameter combinations alternately fixed in each fit.  
Based on collision system and rapidity, variations in \psip counts of 
3.2\%--8.4\% are observed.


The modified Hagedorn function from Eq.~\ref{eq:hag} was also used to 
fit the mixed-events combinatorial background. An uncertainty can be 
introduced in the extracted yields because the mixed-events background 
was normalized over the same mass range used for the like-sign 
background.  The systematic uncertainty was determined following the 
approach described in~\cite{PHENIX:2016vmz}, where the mixed-events 
background was fit over two different mass ranges extending above and 
below the mass range used for the measurements.  
Based on collision system and rapidity, a variation in \psip counts 
of 1.0\%--3.7\% are observed.


The systematic uncertainty associated with fixing the \psip CB 
line shape was calculated by fixing the parameter values with plus/minus 
twice the error in the fit results and repeating the fit.  The extracted 
\psip yields were then compared.  
Based on collision system and rapidity, variations in \psip counts of 
1.0\%--1.4\% are observed.


The CB tail parameters $\alpha$ and $n$ were constrained to 
values determined by simulation.  A toy-model study was used to estimate 
the systematic uncertainty.  The mass spectra randomly generated by the 
toy model are based on high statistics \textsc{Geant4} simulated mass 
histograms as well as the relative yields of \jpsi and \psip observed in 
real \pp collision data~\cite{PHENIX:2016vmz}.  One hundred mass 
distributions were produced, and the resulting yields from fitting the 
toy-model distributions were compared to the input \jpsi and \psip counts to 
gauge the accuracy of the CB tail parameters. The toy model was 
run with $\alpha$ and $n$ fixed to the initial fit result.  The systematic 
was then determined by fixing the parameter values with plus/minus twice 
the error in the fit results and repeating the fits. 
Based on rapidity, variations in \psip counts of 
2.6\%--3.8\% are observed.


A toy model was also used to gauge the accuracy of the second Gaussian 
parameters $f$ and $r$ and the overall fit uncertainty.  Fits to high 
statistics \psip embedded simulations were used to extract the second 
Gaussian parameters.  The toy model generates a new simulated mass 
distribution with each throw that was fit using the same technique as the 
data.  Then the systematic uncertainty was calculated as the percent error 
between the \jpsi and \psip counts input to the toy model and the averaged 
output of the toy model after 100 throws. 
Based on collision system and rapidity, variations in \psip counts of 
2.0\%--4.7\% are observed.
All Type B systematic uncertainties associated with the \psip and \jpsi 
signal extraction are summarized in Tables~\ref{tab:sys_fit1} and 
\ref{tab:sys_fit2}, respectively.

\subsubsection{Acceptance and Efficiency Correction}

Both the trigger efficiency and the combined acceptance and reconstruction 
efficiency were determined using a full \textsc{Geant4} detector 
simulation.  Systematic uncertainties due to differences between data and 
simulation, such as the uncertainty on the MuID efficiency, were 
determined in a previous study~\cite{Acharya:2019zjt}.  The systematic 
uncertainties associated with the efficiency corrections are briefly 
summarized below.


To estimate the run-to-run variation arising from different efficiency 
rates, the data sets were grouped according to instantaneous beam 
luminosity, and the invariant yields were calculated for each group.  A 
variation of 1.6\%--4.0\% was observed depending on rapidity and data set.


The $\varphi$ matching systematic uncertainty due to different detector 
dead areas was estimated by comparing the active MuTr azimuthal angle 
distributions in simulation and data.  A variation of 1.6\%--4.7\% was 
observed depending on rapidity and data set.


The initial-shape systematic includes the uncertainty in the simulated 
dimuon rapidity and \pt distributions, which are tuned to previous PHENIX 
measurements~\cite{PHENIX:2011gyb,PHENIX:2012czk,Adare:2010fn} because these 
distributions are not precisely known.  Two sets of simulated mass 
distributions were compared with different assumptions of the \pt and 
rapidity dependence, and a systematic uncertainty of 2\% is conservatively 
quoted for all collision systems and rapidities.


The dimuon trigger efficiency was determined from simulation using a 
dimuon trigger emulator.  In this approach, a correction is applied to the 
dimuon trigger efficiency based on a comparison of the single-muon trigger 
efficiency obtained in simulation and real data.  A variation of 
1.0\%--4.5\% in systematic uncertainty is observed based on collision 
system and rapidity.
All Type B systematic uncertainties associated with the dimuon acceptance 
and efficiency corrections are summarized in Table~\ref{tab:sys_acceff}.

\begin{table}[htb]
\caption{\label{tab:sys_acceff}
The dimuon fractional systematic uncertainties in \pp, \pal, and \pau
collisions for the combined acceptance and reconstruction efficiency
correction and the trigger efficiency 
correction~\protect\cite{Acharya:2019zjt}.
}
\begin{ruledtabular} \begin{tabular}{ccccc}
Source  & System            & Forward        & Backward        & Type \\
\hline
 Run Variation      & \pp  & 4.0\%   & 4.7\%    & B   \\
                    & \pal & 2.8\%   & 3.3\%    & B   \\
                    & \pau & 1.6\%   & 3.5\%    & B   \\ 
\\
$\varphi$ Matching  & \pp  & 5.8\%   & 5.0\%    & B   \\ 
                    & \pal & 3.6\%   & 3.3\%    & B   \\
                    & \pau & 3.4\%   & 4.0\%    & B   \\ 
\\
 Initial Shape      & all  & 2.0\%   & 2.0\%    & B   \\ 
\\
 Trigger Eff.       & \pp   & 2.3\%     & 2.6\%      & B   \\ 
                    & \pal  & 1.0\%     & 3.0\%      & B   \\
                    & \pau  & 1.0\%--4.5\%  & 1.9\%--3.5\%      & B   
\end{tabular} \end{ruledtabular}
\end{table}

\subsubsection{\meanncoll and BBC Efficiency}

The systematic uncertainty related to the mean number of nucleon-nucleon 
collisions, \meanncoll, was determined by varying the input parameters, 
such as the gold nuclei Woods-Saxon parameters and the inelastic cross 
section of nucleon-nucleon collisions, in a Glauber-model calculation, 
as described in~\cite{PHENIX:2013mpn}. A variation of 6.1\%--7.3\% in 
systematic uncertainty is observed in \pau collisions based on 
centrality, as listed in Table~\ref{tab:centrality}. For 
nuclear-modification measurements as a function of \meanncoll, the 
\meanncoll uncertainty is treated as a Type B systematic uncertainty.  
The Type C systematic uncertainty for the BBC efficiency of 10\% was 
previously determined in \pp collisions at \sqstwo~\cite{Adler:2003pb}.


A previous study of multiple interactions for beam crossing in \pp and 
\pal collisions at $\sqrt{s}=200$~GeV found approximately a 5\% variation 
in the measured invariant yields due to the high instantaneous beam 
luminosity~\cite{Acharya:2019zjt}.  However, the acceptance and efficiency 
correction accounts for the instantaneous beam luminosity dependence, and 
a multiple interactions systematic was not quoted.  The same approach is 
followed here.


PHENIX has recently measured the \jpsi polarization in \pp collisions at 
midrapidity~\cite{PHENIX:2020dqu}, and found the polarization in three 
different frames of reference were all consistent with zero.  
Additionally, LHC experiments~\cite{ALICE:2011gej,LHCb:2013izl, 
CMS:2013gbz} have found no strong evidence in favor of quarkonia 
polarization.  Therefore, zero \psip and \jpsi polarization is assumed, 
and a systematic uncertainty is not included for the measurements 
presented in this paper.
 
 
\section{Results and Discussion}

The \psip nuclear-modification factor as a function of centrality has 
been compared with theoretical predictions with and without 
hot-nuclear-matter effects, provided by H.S. Shao~\textit{et al.} and by 
X. Du and R. Rapp~\cite{Du:2018wsj}.  The gluon-shadowing predictions 
use different parameterizations of the nPDFs, including 
Eskola-Paukkunen-Salgado (EPS09)~\cite{Eskola:2009uj} for the Du~\&~Rapp 
predictions, and coordinated-theoretical-experimental project on QCD 
(nCTEQ15)~\cite{Kovarik:2015cma} and Eskola-Paakkinen-Paukkunen-Salgado 
(EPPS16)~\cite{Eskola:2016oht} for the Shao~\textit{et al.} predictions.

Additionally, to isolate modification due to final-state effects, measuring 
the \psip to \jpsi ratio mostly removes initial-state effects.  The PHENIX 
\psip measurements are presented alongside corresponding measurements from 
the ALICE and LHCb 
Collaborations~\cite{LHCb:2016vqr,LHCb:2013gmv,ALICE:2013snh,ALICE:2014cgk,Acharya:2020rvc}.

\subsection{\psip Results}

\begin{figure}[tbh]  
\includegraphics[width=1.0\linewidth]{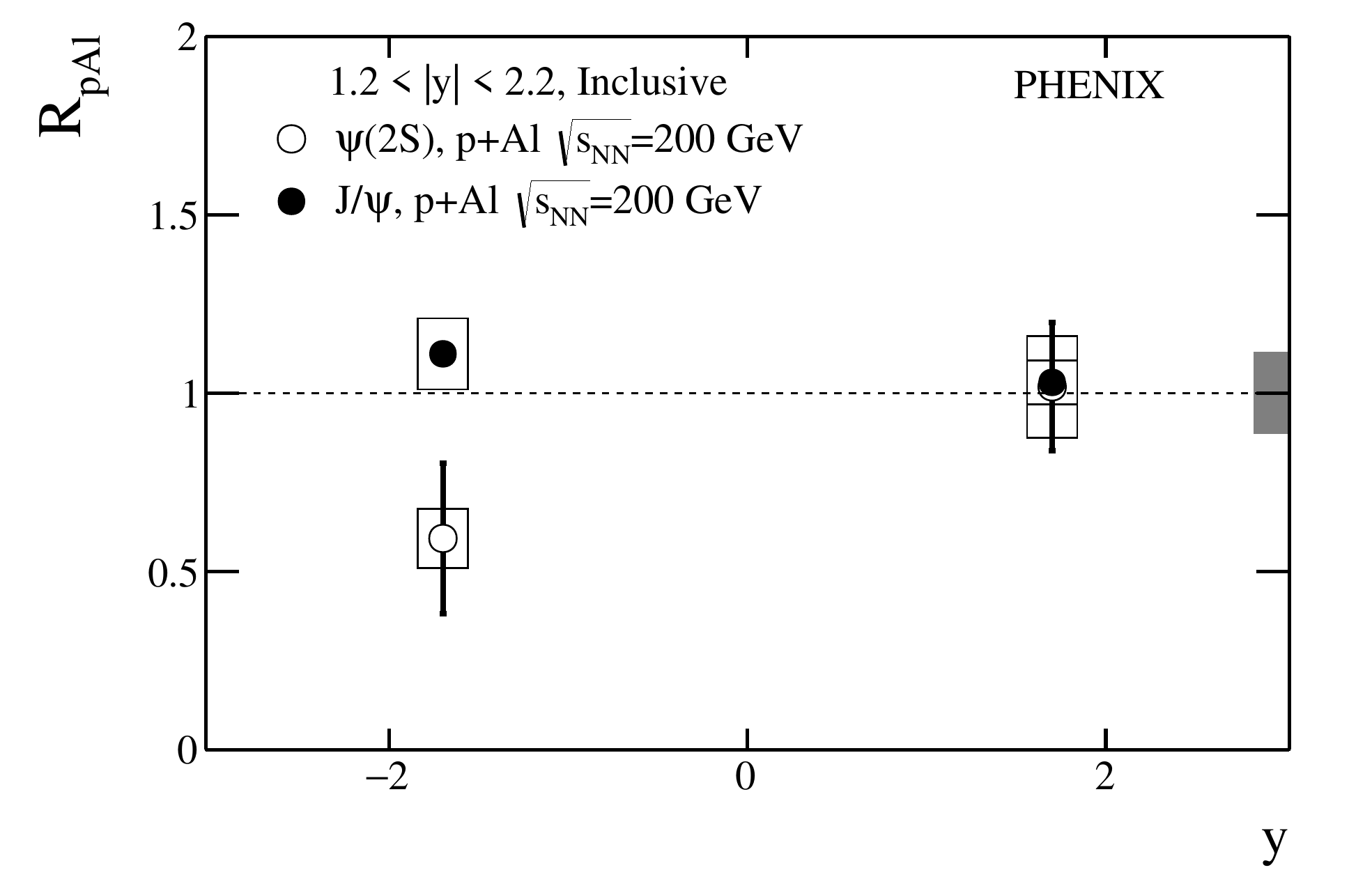}
\caption{\label{fig:RpAl_rap} The centrality and \pt-integrated 
\psip [open circles] and \jpsi [solid circles]
nuclear-modification factors as a function of rapidity in \pal collisions.  
The error boxes (bars) represent point-to-point correlated (uncorrelated) 
uncertainties for each data point.  The Type C global systematic 
uncertainty is indicated by the shaded box on the right side of the 
figure, and includes the BBC efficiency, \meanncoll, and bias correction 
uncertainties.} 
\end{figure}

Figure~\ref{fig:RpAl_rap} shows the \jpsi and \psip nuclear modification 
factors as a function of rapidity in \pal collisions.  At forward 
rapidity, the \jpsi and \psip nuclear modification are consistent with 
unity, as expected based on a weaker gluon modification in the lighter 
Aluminum target system of lower density~\cite{Vogt:2010aa}.  However, at 
backward rapidity, suppression is seen in the \psip modification with 
respect to the \jpsi modification.  Because CNM effects are 
expected to be similar between the two states, nuclear absorption cannot 
explain the suppression at backward rapidity.  The \psip nuclear 
modification results at backward rapidity may indicate final-state effects 
are present in the \pal system at RHIC energies; however, the sizable 
error bars preclude a strong conclusion.  Note that a previous PHENIX 
publication reported a hint of collective flow in 0\%--5\% central \pal 
collisions~\cite{PHENIX:2018hho}.

\begin{figure}[tbh]
\includegraphics[width=1.0\linewidth]{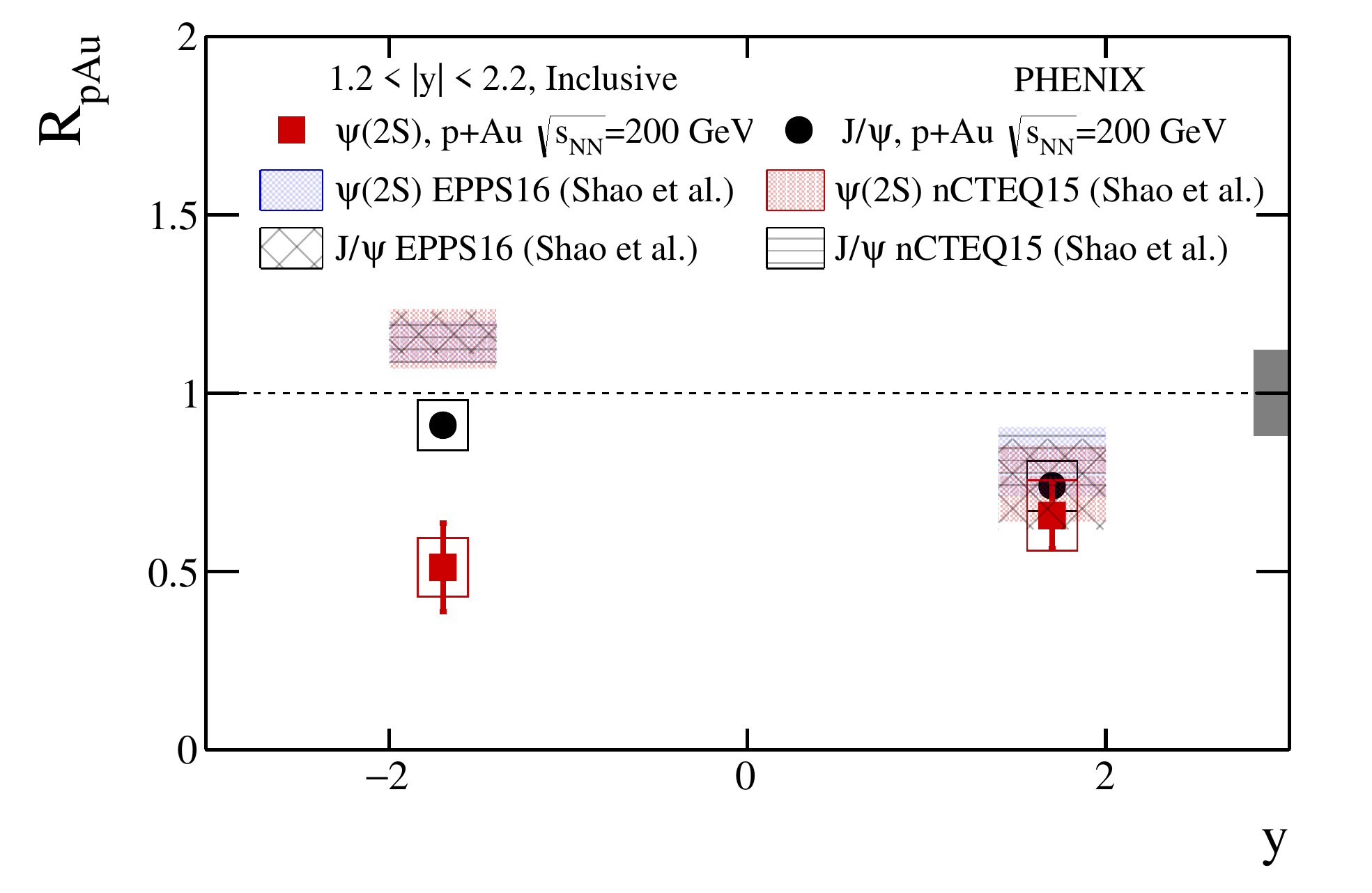}
\caption{\label{fig:RpAu_rap} The centrality and \pt-integrated \psip 
[solid (red) squares] and \jpsi [solid (black) circles] 
nuclear-modification factors as a function of rapidity in \pau 
collisions, compared with EPPS16~\protect\cite{Eskola:2016oht} and 
nCTEQ15~\protect\cite{Kovarik:2015cma} nPDF predictions.  The error 
boxes (bars) represent point-to-point correlated (uncorrelated) 
uncertainties for each data point.  The Type C global systematic 
uncertainty is indicated by the shaded box on the right side of the 
figure, and includes the BBC efficiency, \meanncoll, and bias correction 
uncertainties.  Descriptions of the model predictions are provided in 
the text.}
\end{figure}

Figure~\ref{fig:RpAu_rap} compares the centrality and \pt-integrated 
results for \jpsi and \psip nuclear modification in \pau collisions as a 
function of rapidity with both EPPS16 and nCTEQ15 shadowing predictions 
provided by Shao \textit{et al.}.  The \jpsi EPPS16 (nCTEQ15) 
predictions are also shown. These gluon-shadowing predictions by Shao 
{\it et al.} use a Bayesian-reweighting 
technique~\cite{Kusina:2017gkz,Shao:2012iz,Shao:2015vga,Lansberg:2016deg,Kusina:2020dki} 
for the EPPS16 and nCTEQ15 nPDFs and have a 68\% confidence level.  The 
predictions were calculated at three different factorization scales: 
$0.5\mu_0$, $\mu_0$, and $2\mu_0$, where $\mu_0^2 = M^2 + p_{T}^2$ for 
the quarkonium mass ($M$) and transverse momentum (\pt), and the largest 
of the three uncertainties is quoted.  Previously published PHENIX \jpsi 
and \psip data in \pp collisions~\cite{PHENIX:2011gyb} were fit and used 
as a baseline reference.  The predictions as a function of centrality 
used a Monte-Carlo-Glauber calculation~\cite{Loizides:2017ack} in 
addition to impact-parameter dependent nPDF~\cite{Shao:2020acd}.  The 
modification at forward rapidity is well described by EPPS16 and nCTEQ15 
shadowing, although the \psip nuclear modification shows slightly 
stronger suppression than what shadowing parameterization predicts.  
However, at backward rapidity, both EPPS16 and nCTEQ15 nPDFs overpredict 
the \psip nuclear-modification factor with values of $1.17\pm0.05$ and 
$1.19\pm0.06$, respectively, versus the measured value of $0.51\pm0.12$.  
Therefore, shadowing effects alone cannot describe the \psip suppression 
observed at backward rapidity.

\begin{figure*}[tbh]
\includegraphics[width=1.08\linewidth]{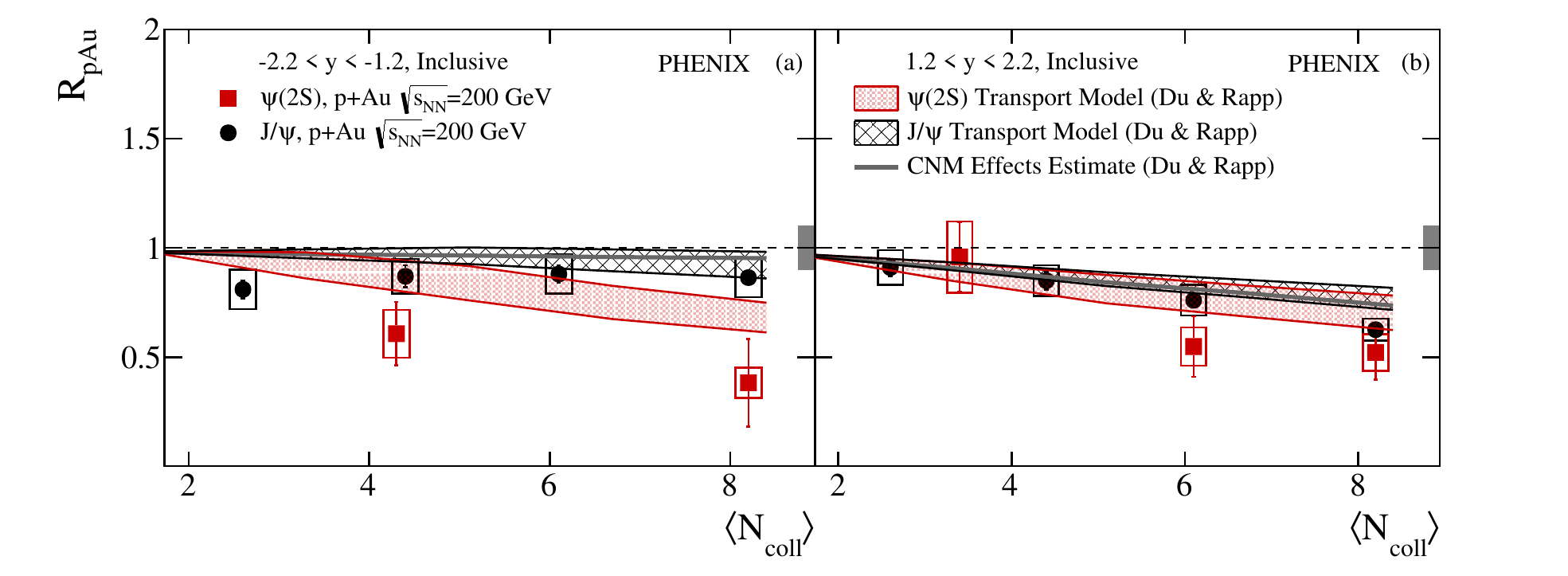}
\caption{\label{fig:RpAu_CNM} The \psip [solid (red) squares] and \jpsi 
[solid (black) circles]~\cite{Acharya:2020rvc} nuclear-modification 
factors as a function of $\langle \Ncoll \rangle$ in \pau collisions at 
(a) backward and (b) forward rapidity is shown with corresponding 
transport-model predictions by Du and Rapp~\protect\cite{Du:2018wsj}.  
Also shown are the transport-model estimates of CNM effects, which are 
the same for both states.  The error boxes (bars) represent 
point-to-point correlated (uncorrelated) uncertainties for each data 
point.  The Type C global systematic uncertainty is indicated by the 
shaded box on the right side of the figure.  Descriptions of the model 
predictions are provided in the text.}
\end{figure*}

\begin{figure*}[tbh]  
\includegraphics[width=1.08\linewidth]{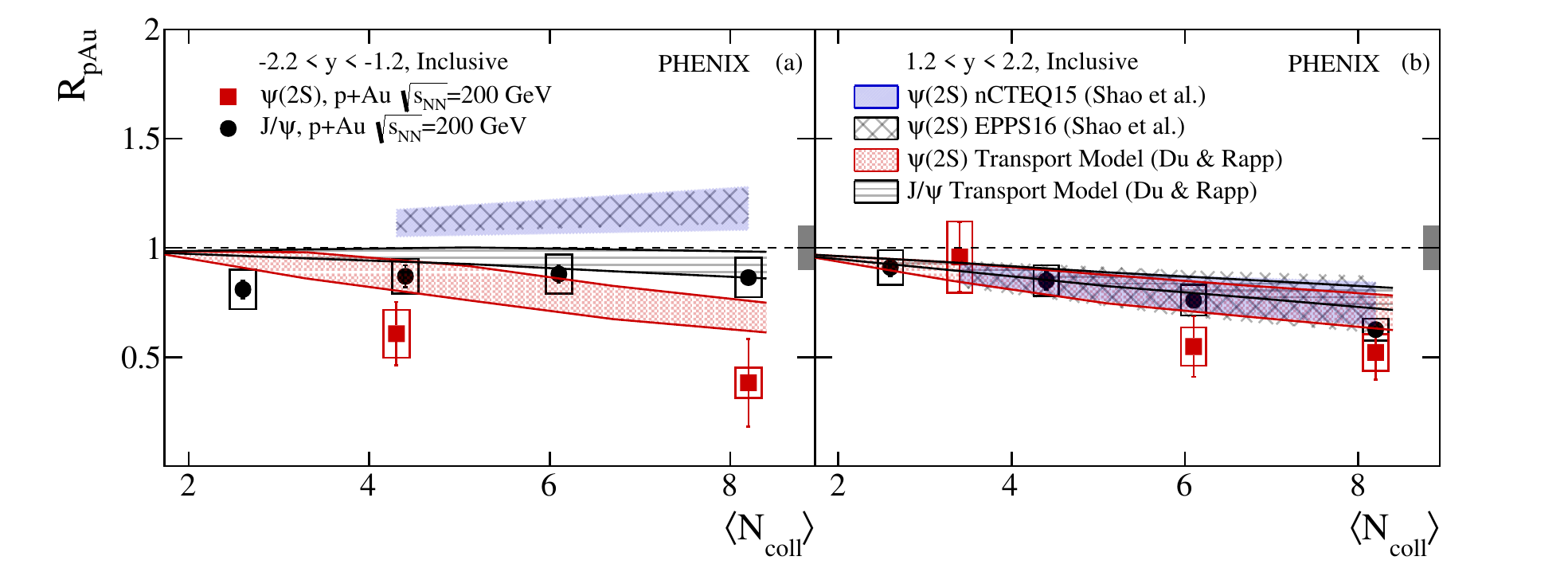}
\caption{\label{fig:Shao_rapp_psip_jpsi} The \psip [solid (red) squares] 
and \jpsi [solid (black) circles]~\protect\cite{Acharya:2020rvc} 
nuclear-modification factors as a function of $\langle \Ncoll \rangle$ 
in \pau collisions at (a) backward and (b) forward rapidity are shown 
with \psip EPPS16~\protect\cite{Eskola:2016oht} and 
nCTEQ15~\protect\cite{Kovarik:2015cma} nPDF predictions.  Also shown are 
the \psip and \jpsi transport-model predictions~\protect\cite{Du:2018wsj}.  
The error boxes (bars) represent point-to-point correlated (uncorrelated) 
uncertainties for each data point.  The Type C global systematic 
uncertainty is indicated by the shaded box on the right side of the 
figure, and includes the BBC efficiency and bias correction 
uncertainties.  See text for descriptions of the model predictions.}
\end{figure*}

\hspace{-0.5cm}
\begin{figure*}[tbh]
\includegraphics[width=1.08\linewidth]{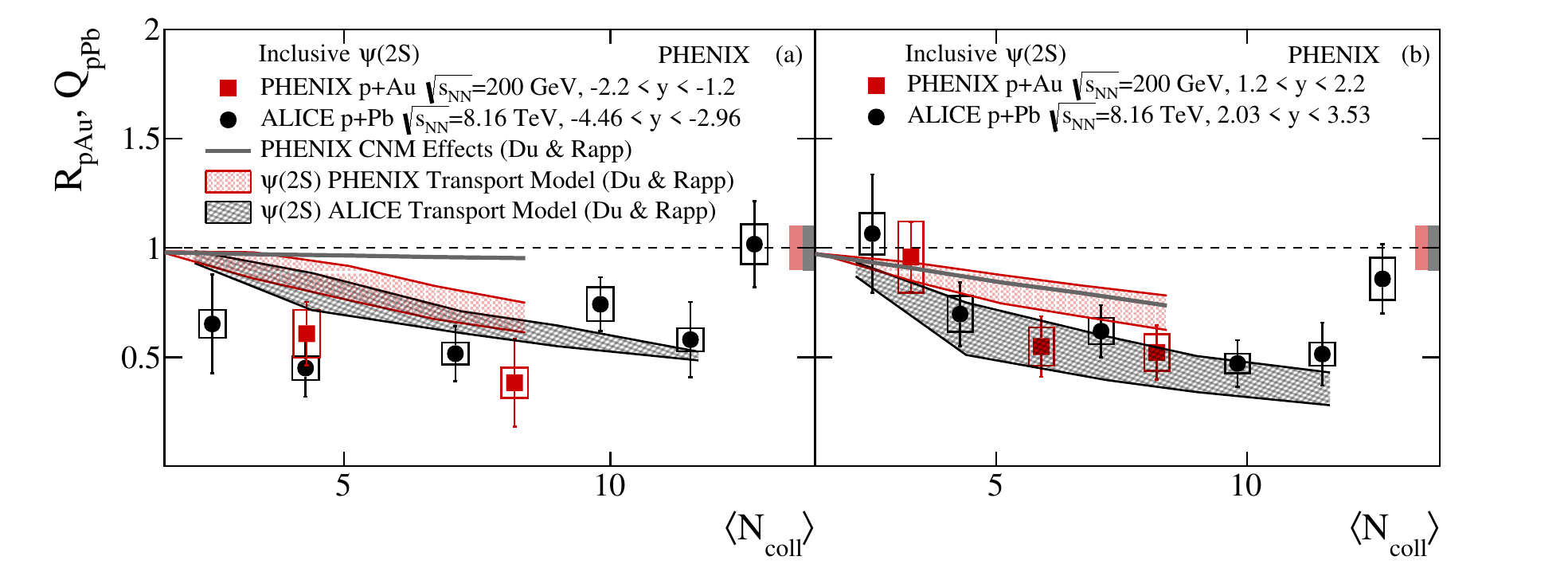}
\caption{\label{fig:RpAu_psip_ncoll_CNM} The PHENIX 
[solid (red) squares] and ALICE [solid (black) circles] 
\psip nuclear-modification factors as a function of $\langle \Ncoll \rangle$ 
in \pau collisions at (a) backward and (b) forward rapidity.
The error boxes (bars) represent point-to-point correlated (uncorrelated) 
uncertainties for each data point.  The Type C global systematic 
uncertainty is indicated by the shaded box on the right side of the 
figure, and includes the BBC efficiency and bias correction 
uncertainties.  Descriptions of the model predictions are provided in 
the text.}
\end{figure*}

Figures~\ref{fig:RpAu_CNM}~and~\ref{fig:Shao_rapp_psip_jpsi} show the 
\jpsi~\cite{Acharya:2020rvc} and \psip nuclear modification measurements 
as a function of \meanncoll.  The nuclear modification between the two 
states follows a similar trend at forward rapidity, with no clear 
difference in suppression in the most central collisions.  The \psip 
shadowing predictions provided by Shao~\textit{et al.} shown in
Fig.~\ref{fig:Shao_rapp_psip_jpsi}(b) underpredict the suppression at 
forward rapidity.  Also shown is a comparison to transport-model (TM) 
predictions for \psip \jpsi provided by Du and Rapp~\cite{Du:2018wsj}.  
The TM was 
originally developed for $A$$+$$A$ collision systems~\cite{Zhao:2010nk}, 
and has been extended to small collision systems.  A nuclear-absorption 
estimate based on the PHENIX \dau data~\cite{PHENIX:2013pmn}, and EPS09 
shadowing effects have been included.  The initial geometry of the 
fireball is derived from a Monte-Carlo-Glauber model.  Gluon shadowing 
from the EPS09 parameterization is the dominant contribution to the TM 
at forward rapidity for both states.  The model reproduces the relative 
suppression, although the degree of suppression is somewhat 
underpredicted.  The \psip TM predicts a small final-state effect, 
which can be seen in Fig.~\ref{fig:RpAu_CNM}(b), where the total
TM predictions and the individual 
contribution from CNM effects alone are shown.

A difference can be seen at backward rapidity between the \jpsi and \psip 
nuclear modification, consistent with a 2.9$\sigma$ effect.  The \psip 
antishadowing predictions provided by Shao \textit{et al.} shown in 
Fig.~\ref{fig:Shao_rapp_psip_jpsi}(a) do not predict the 
suppression.  These predictions are purely antishadowing and do not 
contain any additional CNM effects, such as nuclear 
absorption.  The measured \jpsi nuclear modification is nearly constant as 
a function of centrality.  This behavior may arise as a tradeoff between 
the competing effects of nuclear-thickness-dependent antishadowing 
enhancement and nuclear-absorption suppression~\cite{Acharya:2019zjt, 
McGlinchey:2012bp}.  Figure~\ref{fig:Shao_rapp_psip_jpsi}(a) also shows 
a comparison to TM predictions at backward rapidity, and 
final-state effects and nuclear absorption are expected to be important.  
In addition to gluon antishadowing predictions, the TM  
includes a nuclear-absorption estimate for both the \psip and \jpsi.  
The TMs underpredict the suppression, but describe the 
relative modification well and indicate that the \psip suppression in 
\pau collisions is consistent with final-state effects.  The individual 
contributions from CNM effects alone can be seen in 
Fig.~\ref{fig:RpAu_CNM}, which shows that hot-nuclear-matter effects are 
the predominant source of the stronger \psip suppression at backward 
rapidity.


Figure~\ref{fig:RpAu_psip_ncoll_CNM} shows the PHENIX \psip and the 
ALICE \psip nuclear-modification factors~\cite{Acharya:2020rvc} as a 
function of \meanncoll.  Please refer to the ALICE 
publication~\cite{Acharya:2020rvc} for more information regarding the 
notation $Q_{pPb}$ used for the nuclear-modification factor 
measurements.  Due to the higher energy at the LHC, the nucleon-nucleon 
cross section is larger, which leads to a larger range of \meanncoll 
values for the ALICE data.  The Bjorken-$x$ values probed in the target 
at LHC energies are smaller than the values probed at RHIC energies.  
Additionally, the $Q^2$ values are higher at LHC energies due to the 
larger mean \pt values.  The Du and Rapp TM 
predictions~\cite{Du:2018wsj} are compared with the experimental data. 
At backward rapidity where hot-nuclear-matter effects are dominant, both 
TMs predict a similar degree of suppression.

\begin{figure*}[tbh]
\includegraphics[width=1.08\linewidth]{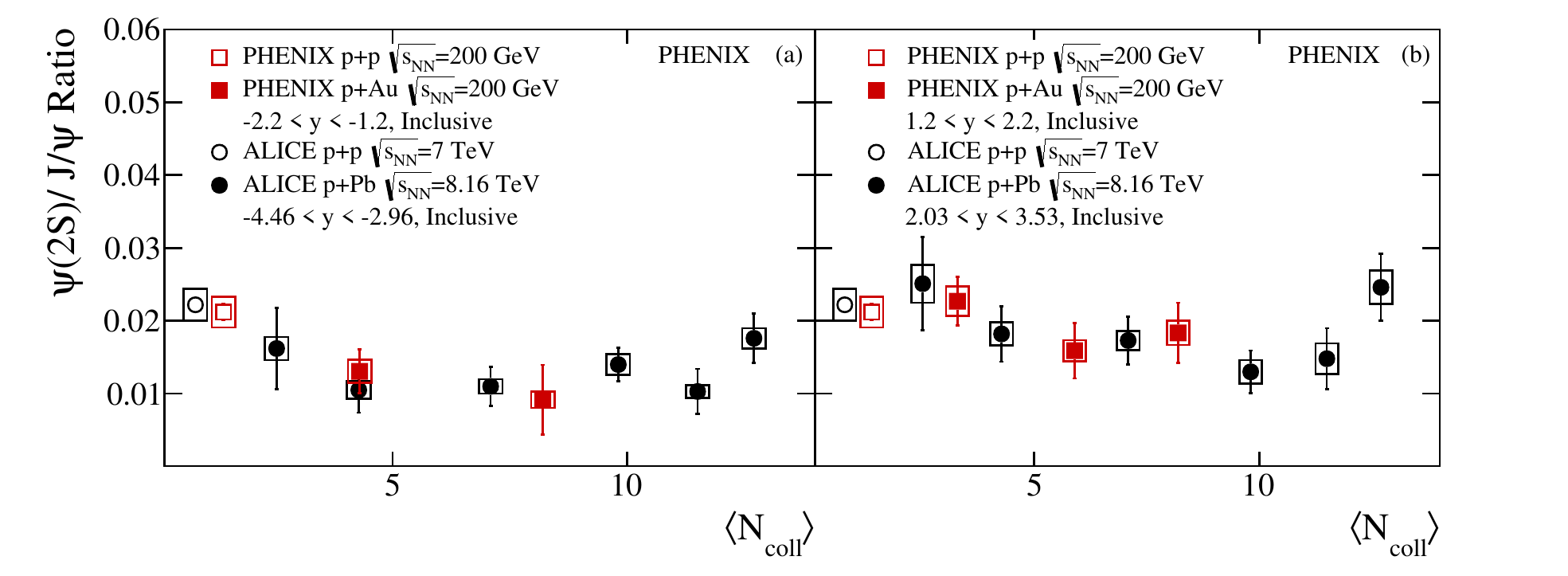}
\caption{\label{fig:ratio_ncoll} 
The PHENIX [solid (red) squares] \psip to \jpsi ratios as a function of 
$\langle \Ncoll \rangle$ in \pau collisions at (a) backward and (b) 
forward rapidity is compared with ALICE [solid (black) circles] 
ratios in $p$ $+$Pb collisions~\cite{Acharya:2020rvc}.  The 
data points at $\langle \Ncoll \rangle$=1 are the equivalent PHENIX 
[open (red) squares] and ALICE~\cite{ALICE:2014uja} [open (black) 
circles] ratios in \pp collisions.  The error boxes (bars) represent 
point-to-point correlated (uncorrelated) uncertainties for each data 
point, and includes the BBC efficiency and bias correction uncertainties.}
\end{figure*}

Figure~\ref{fig:ratio_ncoll} compares the \psip to \jpsi ratios for 
PHENIX and ALICE~\cite{Acharya:2020rvc} as a function of \meanncoll.  
The PHENIX \psip to \jpsi ratio in \pp collisions is shown at $\langle 
N_{\rm coll} \rangle$ = 1.  By taking the ratio, any initial state 
effects are expected to largely cancel because contributions to the 
charmonium modification should be similar between the two states.  At 
backward rapidity, a stronger suppression is seen for the \psip with 
respect to the \jpsi relative to forward rapidity, which can be observed 
in comparison with the \pp reference measurement.  The observed decrease 
of the \pA ratio with respect to the ratio in \pp collisions strongly 
suggests the presence of final-state effects in \pA collisions.  Perhaps 
surprisingly, the comparison of PHENIX and ALICE \psip to \jpsi ratios 
indicates that final-state effects at RHIC and LHC energies are similar.  
This could be due to a combination of hotter temperature and longer 
lifetime of quark-gluon plasma forming at LHC energies and a shorter 
exposure to the medium due to higher mean \pt. Note that in \pp 
collisions, the \psip to \jpsi ratio from world data shows no clear 
energy dependence as a function of center of mass 
energy~\cite{PHENIX:2016vmz}.

\begin{figure}[tbh]
\includegraphics[width=1.0\linewidth]{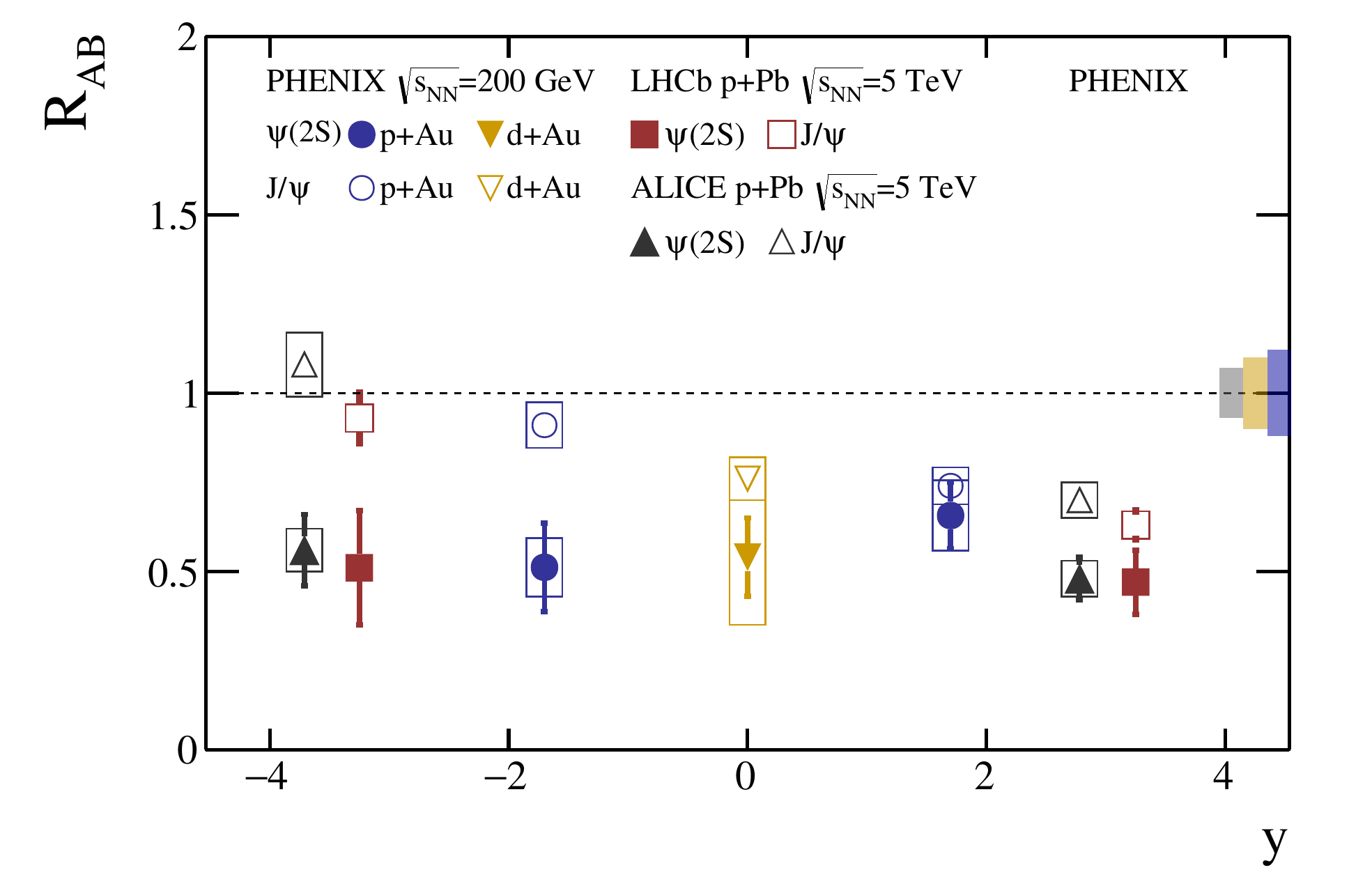}
\caption{\label{fig:RAB_rap} 
The centrality- and \pt-integrated \jpsi and \psip nuclear-modification 
factors as a function of rapidity are shown for small systems at PHENIX 
[downward (gold) triangles~\protect\cite{PHENIX:2013pmn} and (blue) 
circles], LHCb [(red) squares~\protect\cite{LHCb:2016vqr,LHCb:2013gmv}], 
and ALICE [upward (black) 
triangles~\protect\cite{ALICE:2013snh,ALICE:2014cgk}].  The error boxes 
(bars) for the measurements represent point-to-point correlated 
(uncorrelated) uncertainties for each data point.  The Type C global 
systematic uncertainty is indicated by the shaded box on the right side 
of the figure, and includes the BBC efficiency, \meanncoll, and bias 
correction uncertainties.}
\end{figure}

Figure~\ref{fig:RAB_rap} presents measurements of the \jpsi and \psip 
nuclear-modification factors in small-system collision systems from 
three different experimental collaborations.  The PHENIX measurements 
are shown for \dau collisions~\cite{PHENIX:2013pmn} and \pau collisions.  
The LHCb~\cite{LHCb:2016vqr,LHCb:2013gmv} and 
ALICE~\cite{ALICE:2013snh,ALICE:2014cgk} measurements are shown for \ppb 
collisions.  At forward rapidity, the \jpsi and \psip suppression are 
similar, suggesting that initial-state effects dominate charmonium 
modification.  At backward rapidity, the results show a larger 
differential suppression.  The nuclear-modification results from PHENIX, 
LHCb, and ALICE are consistent with increasing final-state effects in 
small systems for rapidities in the $A$-going direction.
 
 
\section{Summary and Conclusions}


The results presented in this paper address initial- and final-state 
effects on charmonium production in \pp, \pal, and \pau collisions at 
\sqsntwo.  The \jpsi and \psip nuclear-modification results and the 
measured \psip to \jpsi ratios are shown for \pau collisions as a 
function of \meanncoll. The \pt- and centrality-integrated \jpsi and 
\psip nuclear-modification factors in \pau and \pal collisions were 
also presented as a function of rapidity.

At forward rapidity, the re-weighted EPPS16~\cite{Eskola:2016oht} and 
nCTEQ15~\cite{Kovarik:2015cma} nPDF predictions predictions slightly 
underestimate the suppression of the Psi(2S) nuclear-modification 
measurements.  Initial-state effects are expected to produce similar 
modification for the \jpsi and \psip states.  The data show the \psip 
suppression is similar to the \jpsi suppression in the most central 
collisions at forward rapidity, suggesting the dominant contribution to 
nuclear-modification is gluon shadowing.  At backward rapidity, the 
antishadowing predictions alone cannot reproduce the \psip 
nuclear-modification data and suggests additional effects beyond gluon 
antishadowing are present.

In a previous PHENIX publication~\cite{Acharya:2019zjt}, it was shown that 
the centrality-dependent suppression seen in the \jpsi nuclear 
modification in \pau collisions at backward rapidity was consistent with a 
tradeoff between gluon-antishadowing and nuclear absorption.  There will 
also be a tradeoff in the \pal system between absorption and 
antishadowing, although both will be weaker.  No suppression was observed 
at backward rapidity for the \jpsi nuclear modification in \pal 
collisions.  The difference in suppression between the \jpsi and \psip 
states is suggestive of final-state effects, although a strong statement 
is not possible due to the large experimental uncertainties.


Comparing the TM predictions with the data, the relative 
suppression between \jpsi and \psip is well described, although the 
overall suppression is underestimated. As seen in 
Fig.~\ref{fig:RpAu_CNM}(b), the model predicts slightly stronger 
suppression for the \psip at forward rapidity due to a small final-state 
effect. Also seen in Fig.~\ref{fig:RpAu_CNM}(a), at 
backward rapidity the model predicts significantly more suppression of the 
\psip than the \jpsi.  This is due to a stronger final-state effect for 
the \psip because contributions from CNM effects are the 
same in both the \psip and \jpsi TM predictions.


At backward rapidity, the PHENIX and ALICE \psip nuclear modification 
results are surprisingly similar.  The TM calculations at the 
two energies reproduce this well.  The PHENIX \psip to \jpsi ratio was 
compared with the ALICE ratio as a function of \meanncoll to cancel most 
of the modifications due to initial-state effects, which are expected to 
be different at the two energies.  The PHENIX and ALICE \psip to \jpsi 
ratio measurements follow the same trend at backward rapidity, indicating 
that final-state effects on inclusive charmonium states appears to be very 
similar at RHIC and LHC energies.


In comparing PHENIX \jpsi and \psip nuclear modification with LHCb and 
ALICE results, at forward rapidity the \psip nuclear modification is 
slightly more suppressed than the \jpsi nuclear modification at the most 
forward rapidity, suggesting that initial-state effects are the dominant 
contribution. At backward rapidity, a clear trend is seen where the \psip 
is more suppressed than the \jpsi.  This observed behavior reported by 
three different experiments is consistent with transport models that 
include hot-nuclear-matter effects in the $A$-going direction.
 


\begin{acknowledgments}

We thank the staff of the Collider-Accelerator and Physics
Departments at Brookhaven National Laboratory and the staff of
the other PHENIX participating institutions for their vital
contributions.  
We acknowledge support from the Office of Nuclear Physics in the
Office of Science of the Department of Energy,
the National Science Foundation,
Abilene Christian University Research Council,
Research Foundation of SUNY, and
Dean of the College of Arts and Sciences, Vanderbilt University
(U.S.A),
Ministry of Education, Culture, Sports, Science, and Technology
and the Japan Society for the Promotion of Science (Japan),
Natural Science Foundation of China (People's Republic of China),
Croatian Science Foundation and
Ministry of Science and Education (Croatia),
Ministry of Education, Youth and Sports (Czech Republic),
Centre National de la Recherche Scientifique, Commissariat
{\`a} l'{\'E}nergie Atomique, and Institut National de Physique
Nucl{\'e}aire et de Physique des Particules (France),
J. Bolyai Research Scholarship, EFOP, the New National Excellence
Program ({\'U}NKP), NKFIH, and OTKA (Hungary),
Department of Atomic Energy and Department of Science and Technology
(India),
Israel Science Foundation (Israel),
Basic Science Research and SRC(CENuM) Programs through NRF
funded by the Ministry of Education and the Ministry of
Science and ICT (Korea).
Ministry of Education and Science, Russian Academy of Sciences,
Federal Agency of Atomic Energy (Russia),
VR and Wallenberg Foundation (Sweden),
University of Zambia, the Government of the Republic of Zambia (Zambia),
the U.S. Civilian Research and Development Foundation for the
Independent States of the Former Soviet Union,
the Hungarian American Enterprise Scholarship Fund,
the US-Hungarian Fulbright Foundation,
and the US-Israel Binational Science Foundation.

\end{acknowledgments}
 

\begin{thebibliography}{65}%
\makeatletter
\providecommand \@ifxundefined [1]{%
 \@ifx{#1\undefined}
}%
\providecommand \@ifnum [1]{%
 \ifnum #1\expandafter \@firstoftwo
 \else \expandafter \@secondoftwo
 \fi
}%
\providecommand \@ifx [1]{%
 \ifx #1\expandafter \@firstoftwo
 \else \expandafter \@secondoftwo
 \fi
}%
\providecommand \natexlab [1]{#1}%
\providecommand \enquote  [1]{``#1''}%
\providecommand \bibnamefont  [1]{#1}%
\providecommand \bibfnamefont [1]{#1}%
\providecommand \citenamefont [1]{#1}%
\providecommand \href@noop [0]{\@secondoftwo}%
\providecommand \href [0]{\begingroup \@sanitize@url \@href}%
\providecommand \@href[1]{\@@startlink{#1}\@@href}%
\providecommand \@@href[1]{\endgroup#1\@@endlink}%
\providecommand \@sanitize@url [0]{\catcode `\\12\catcode `\$12\catcode
  `\&12\catcode `\#12\catcode `\^12\catcode `\_12\catcode `\%12\relax}%
\providecommand \@@startlink[1]{}%
\providecommand \@@endlink[0]{}%
\providecommand \url  [0]{\begingroup\@sanitize@url \@url }%
\providecommand \@url [1]{\endgroup\@href {#1}{\urlprefix }}%
\providecommand \urlprefix  [0]{URL }%
\providecommand \Eprint [0]{\href }%
\providecommand \doibase [0]{https://doi.org/}%
\providecommand \selectlanguage [0]{\@gobble}%
\providecommand \bibinfo  [0]{\@secondoftwo}%
\providecommand \bibfield  [0]{\@secondoftwo}%
\providecommand \translation [1]{[#1]}%
\providecommand \BibitemOpen [0]{}%
\providecommand \bibitemStop [0]{}%
\providecommand \bibitemNoStop [0]{.\EOS\space}%
\providecommand \EOS [0]{\spacefactor3000\relax}%
\providecommand \BibitemShut  [1]{\csname bibitem#1\endcsname}%
\let\auto@bib@innerbib\@empty
\bibitem [{\citenamefont {Nagle}\ and\ \citenamefont
  {Zajc}(2018)}]{Nagle:2018nvi}%
  \BibitemOpen
  \bibfield  {author} {\bibinfo {author} {\bibfnamefont {J.~L.}\ \bibnamefont
  {Nagle}}\ and\ \bibinfo {author} {\bibfnamefont {W.~A.}\ \bibnamefont
  {Zajc}},\ }\bibfield  {title} {\bibinfo {title} {{Small System Collectivity
  in Relativistic Hadronic and Nuclear Collisions}},\ }\href
  {https://doi.org/10.1146/annurev-nucl-101916-123209} {\bibfield  {journal}
  {\bibinfo  {journal} {Ann. Rev. Nucl. Part. Sci.}\ }\textbf {\bibinfo
  {volume} {68}},\ \bibinfo {pages} {211} (\bibinfo {year} {2018})}\BibitemShut
  {NoStop}%
\bibitem [{\citenamefont {Khachatryan}\ \emph {et~al.}()\citenamefont
  {Khachatryan} \emph {et~al.}}]{CMS:2010ifv}%
  \BibitemOpen
  \bibfield  {author} {\bibinfo {author} {\bibfnamefont {V.}~\bibnamefont
  {Khachatryan}} \emph {et~al.} (\bibinfo {collaboration} {CMS
  Collaboration}),\ }\bibfield  {title} {\bibinfo {title} {{Observation of
  Long-Range Near-Side Angular Correlations in Proton-Proton Collisions at the
  LHC}},\ }\bibinfo {note} {{J. High Energy Phys. {\bf 09 (2010)},
  091}}\BibitemShut {NoStop}%
\bibitem [{\citenamefont {Abelev}\ \emph {et~al.}(2013)\citenamefont {Abelev}
  \emph {et~al.}}]{ALICE:2012eyl}%
  \BibitemOpen
  \bibfield  {author} {\bibinfo {author} {\bibfnamefont {B.}~\bibnamefont
  {Abelev}} \emph {et~al.} (\bibinfo {collaboration} {ALICE Collaboration}),\
  }\bibfield  {title} {\bibinfo {title} {{Long-range angular correlations on
  the near and away side in $p$-Pb collisions at $\sqrt{s_{NN}}=5.02$ TeV}},\
  }\href {https://doi.org/10.1016/j.physletb.2013.01.012} {\bibfield  {journal}
  {\bibinfo  {journal} {Phys. Lett. B}\ }\textbf {\bibinfo {volume} {719}},\
  \bibinfo {pages} {29} (\bibinfo {year} {2013})}\BibitemShut {NoStop}%
\bibitem [{\citenamefont {Adamczyk}\ \emph
  {et~al.}(2016{\natexlab{a}})\citenamefont {Adamczyk} \emph
  {et~al.}}]{STAR:2016ydv}%
  \BibitemOpen
  \bibfield  {author} {\bibinfo {author} {\bibfnamefont {L.}~\bibnamefont
  {Adamczyk}} \emph {et~al.} (\bibinfo {collaboration} {STAR Collaboration}),\
  }\bibfield  {title} {\bibinfo {title} {{Measurement of elliptic flow of light
  nuclei at $\sqrt{s_{NN}}=$ 200, 62.4, 39, 27, 19.6, 11.5, and 7.7 GeV at the
  BNL Relativistic Heavy Ion Collider}},\ }\href
  {https://doi.org/10.1103/PhysRevC.94.034908} {\bibfield  {journal} {\bibinfo
  {journal} {Phys. Rev. C}\ }\textbf {\bibinfo {volume} {94}},\ \bibinfo
  {pages} {034908} (\bibinfo {year} {2016}{\natexlab{a}})}\BibitemShut
  {NoStop}%
\bibitem [{\citenamefont {Aaboud}\ \emph {et~al.}(2017)\citenamefont {Aaboud}
  \emph {et~al.}}]{ATLAS:2017hap}%
  \BibitemOpen
  \bibfield  {author} {\bibinfo {author} {\bibfnamefont {M.}~\bibnamefont
  {Aaboud}} \emph {et~al.} (\bibinfo {collaboration} {ATLAS Collaboration}),\
  }\bibfield  {title} {\bibinfo {title} {{Measurement of multi-particle
  azimuthal correlations in $pp$, $p$$+$Pb and low-multiplicity Pb$+$Pb
  collisions with the ATLAS detector}},\ }\href
  {https://doi.org/10.1140/epjc/s10052-017-4988-1} {\bibfield  {journal}
  {\bibinfo  {journal} {Eur. Phys. J. C}\ }\textbf {\bibinfo {volume} {77}},\
  \bibinfo {pages} {428} (\bibinfo {year} {2017})}\BibitemShut {NoStop}%
\bibitem [{\citenamefont {Aidala}\ \emph
  {et~al.}(2019{\natexlab{a}})\citenamefont {Aidala} \emph
  {et~al.}}]{PHENIX:2018lia}%
  \BibitemOpen
  \bibfield  {author} {\bibinfo {author} {\bibfnamefont {C.}~\bibnamefont
  {Aidala}} \emph {et~al.} (\bibinfo {collaboration} {PHENIX Collaboration}),\
  }\bibfield  {title} {\bibinfo {title} {{Creation of quark-gluon plasma
  droplets with three distinct geometries}},\ }\href
  {https://doi.org/10.1038/s41567-018-0360-0} {\bibfield  {journal} {\bibinfo
  {journal} {Nature Phys.}\ }\textbf {\bibinfo {volume} {15}},\ \bibinfo
  {pages} {214} (\bibinfo {year} {2019}{\natexlab{a}})}\BibitemShut {NoStop}%
\bibitem [{\citenamefont {Acharya}\ \emph {et~al.}(2022)\citenamefont {Acharya}
  \emph {et~al.}}]{PHENIX:2021bxz}%
  \BibitemOpen
  \bibfield  {author} {\bibinfo {author} {\bibfnamefont {U.~A.}\ \bibnamefont
  {Acharya}} \emph {et~al.} (\bibinfo {collaboration} {PHENIX Collaboration}),\
  }\bibfield  {title} {\bibinfo {title} {{Kinematic dependence of azimuthal
  anisotropies in $p$$+$Au, $d$$+$Au, $^3$He$+$Au at $\sqrt{s_{_{NN}}}=200$
  GeV}},\ }\href {https://doi.org/10.1103/PhysRevC.105.024901} {\bibfield
  {journal} {\bibinfo  {journal} {Phys. Rev. C}\ }\textbf {\bibinfo {volume}
  {105}},\ \bibinfo {pages} {024901} (\bibinfo {year} {2022})}\BibitemShut
  {NoStop}%
\bibitem [{\citenamefont {Adare}\ \emph
  {et~al.}(2013{\natexlab{a}})\citenamefont {Adare} \emph
  {et~al.}}]{PHENIX:2013pmn}%
  \BibitemOpen
  \bibfield  {author} {\bibinfo {author} {\bibfnamefont {A.}~\bibnamefont
  {Adare}} \emph {et~al.} (\bibinfo {collaboration} {PHENIX Collaboration}),\
  }\bibfield  {title} {\bibinfo {title} {{Nuclear Modification of $\psi\prime$,
  $\chi_{c}$, and J/\ensuremath{\psi} Production in $d$+Au Collisions at
  $\sqrt{s_{NN}}$=200 GeV}},\ }\href
  {https://doi.org/10.1103/PhysRevLett.111.202301} {\bibfield  {journal}
  {\bibinfo  {journal} {Phys. Rev. Lett.}\ }\textbf {\bibinfo {volume} {111}},\
  \bibinfo {pages} {202301} (\bibinfo {year} {2013}{\natexlab{a}})}\BibitemShut
  {NoStop}%
\bibitem [{\citenamefont {Abelev}\ \emph {et~al.}({\natexlab{a}})\citenamefont
  {Abelev} \emph {et~al.}}]{ALICE:2014cgk}%
  \BibitemOpen
  \bibfield  {author} {\bibinfo {author} {\bibfnamefont {B.~B.}\ \bibnamefont
  {Abelev}} \emph {et~al.} (\bibinfo {collaboration} {ALICE Collaboration}),\
  }\bibfield  {title} {\bibinfo {title} {{Suppression of $\psi$(2S) production
  in p-Pb collisions at $\sqrt{s_{\rm NN}}$ = 5.02 TeV}}} ({\natexlab{a}}),\
  \bibinfo {note} {{J. High Energy Phys. {\bf 12 (2014)}, 073}}\BibitemShut
  {NoStop}%
\bibitem [{\citenamefont {Aaij}\ \emph {et~al.}({\natexlab{a}})\citenamefont
  {Aaij} \emph {et~al.}}]{LHCb:2016vqr}%
  \BibitemOpen
  \bibfield  {author} {\bibinfo {author} {\bibfnamefont {R.}~\bibnamefont
  {Aaij}} \emph {et~al.} (\bibinfo {collaboration} {LHCb Collaboration}),\
  }\bibfield  {title} {\bibinfo {title} {{Study of $\psi(2S)$ production and
  cold nuclear matter effects in $p$Pb collisions at
  $\sqrt{s_{NN}}=5~\mathrm{TeV}$}}} ({\natexlab{a}}),\ \bibinfo {note} {{J.
  High Energy Phys. {\bf 03 (2016)}, 133}}\BibitemShut {NoStop}%
\bibitem [{\citenamefont {Adare}\ \emph
  {et~al.}(2013{\natexlab{b}})\citenamefont {Adare} \emph
  {et~al.}}]{PHENIX:2012czk}%
  \BibitemOpen
  \bibfield  {author} {\bibinfo {author} {\bibfnamefont {A.}~\bibnamefont
  {Adare}} \emph {et~al.} (\bibinfo {collaboration} {PHENIX Collaboration}),\
  }\bibfield  {title} {\bibinfo {title} {{Transverse-Momentum Dependence of the
  $J/\psi$ Nuclear Modification in $d+$Au Collisions at $\sqrt{s_{NN}}=200$
  GeV}},\ }\href {https://doi.org/10.1103/PhysRevC.87.034904} {\bibfield
  {journal} {\bibinfo  {journal} {Phys. Rev. C}\ }\textbf {\bibinfo {volume}
  {87}},\ \bibinfo {pages} {034904} (\bibinfo {year}
  {2013}{\natexlab{b}})}\BibitemShut {NoStop}%
\bibitem [{\citenamefont {Brambilla}\ \emph {et~al.}(2011)\citenamefont
  {Brambilla} \emph {et~al.}}]{Brambilla:2010cs}%
  \BibitemOpen
  \bibfield  {author} {\bibinfo {author} {\bibfnamefont {N.}~\bibnamefont
  {Brambilla}} \emph {et~al.},\ }\bibfield  {title} {\bibinfo {title} {{Heavy
  Quarkonium: Progress, Puzzles, and Opportunities}},\ }\href
  {https://doi.org/10.1140/epjc/s10052-010-1534-9} {\bibfield  {journal}
  {\bibinfo  {journal} {Eur. Phys. J. C}\ }\textbf {\bibinfo {volume} {71}},\
  \bibinfo {pages} {1534} (\bibinfo {year} {2011})}\BibitemShut {NoStop}%
\bibitem [{\citenamefont {Eskola}\ \emph {et~al.}(2017)\citenamefont {Eskola},
  \citenamefont {Paakkinen}, \citenamefont {Paukkunen},\ and\ \citenamefont
  {Salgado}}]{Eskola:2016oht}%
  \BibitemOpen
  \bibfield  {author} {\bibinfo {author} {\bibfnamefont {K.~J.}\ \bibnamefont
  {Eskola}}, \bibinfo {author} {\bibfnamefont {P.}~\bibnamefont {Paakkinen}},
  \bibinfo {author} {\bibfnamefont {H.}~\bibnamefont {Paukkunen}},\ and\
  \bibinfo {author} {\bibfnamefont {C.~A.}\ \bibnamefont {Salgado}},\
  }\bibfield  {title} {\bibinfo {title} {{EPPS16: Nuclear parton distributions
  with LHC data}},\ }\href {https://doi.org/10.1140/epjc/s10052-017-4725-9}
  {\bibfield  {journal} {\bibinfo  {journal} {Eur. Phys. J. C}\ }\textbf
  {\bibinfo {volume} {77}},\ \bibinfo {pages} {163} (\bibinfo {year}
  {2017})}\BibitemShut {NoStop}%
\bibitem [{\citenamefont {Kovarik}\ \emph {et~al.}(2016)\citenamefont {Kovarik}
  \emph {et~al.}}]{Kovarik:2015cma}%
  \BibitemOpen
  \bibfield  {author} {\bibinfo {author} {\bibfnamefont {K.}~\bibnamefont
  {Kovarik}} \emph {et~al.},\ }\bibfield  {title} {\bibinfo {title} {{nCTEQ15-
  Global analysis of nuclear parton distributions with uncertainties in the
  CTEQ framework}},\ }\href {https://doi.org/10.1103/PhysRevD.93.085037}
  {\bibfield  {journal} {\bibinfo  {journal} {Phys. Rev. D}\ }\textbf {\bibinfo
  {volume} {93}},\ \bibinfo {pages} {085037} (\bibinfo {year}
  {2016})}\BibitemShut {NoStop}%
\bibitem [{\citenamefont {McGlinchey}\ \emph {et~al.}(2013)\citenamefont
  {McGlinchey}, \citenamefont {Frawley},\ and\ \citenamefont
  {Vogt}}]{McGlinchey:2012bp}%
  \BibitemOpen
  \bibfield  {author} {\bibinfo {author} {\bibfnamefont {D.~C.}\ \bibnamefont
  {McGlinchey}}, \bibinfo {author} {\bibfnamefont {A.~D.}\ \bibnamefont
  {Frawley}},\ and\ \bibinfo {author} {\bibfnamefont {R.}~\bibnamefont
  {Vogt}},\ }\bibfield  {title} {\bibinfo {title} {{Impact parameter dependence
  of the nuclear modification of $J/\psi$ production in $d+$Au collisions at
  $\sqrt{s_{NN}} = 200$ GeV}},\ }\href
  {https://doi.org/10.1103/PhysRevC.87.054910} {\bibfield  {journal} {\bibinfo
  {journal} {Phys. Rev. C}\ }\textbf {\bibinfo {volume} {87}},\ \bibinfo
  {pages} {054910} (\bibinfo {year} {2013})}\BibitemShut {NoStop}%
\bibitem [{\citenamefont {Arleo}\ \emph {et~al.}(2000)\citenamefont {Arleo},
  \citenamefont {Gossiaux}, \citenamefont {Gousset},\ and\ \citenamefont
  {Aichelin}}]{Arleo:1999af}%
  \BibitemOpen
  \bibfield  {author} {\bibinfo {author} {\bibfnamefont {F.}~\bibnamefont
  {Arleo}}, \bibinfo {author} {\bibfnamefont {P.~B.}\ \bibnamefont {Gossiaux}},
  \bibinfo {author} {\bibfnamefont {T.}~\bibnamefont {Gousset}},\ and\ \bibinfo
  {author} {\bibfnamefont {J.}~\bibnamefont {Aichelin}},\ }\bibfield  {title}
  {\bibinfo {title} {{Charmonium suppression in $pA$ collisions}},\ }\href
  {https://doi.org/10.1103/PhysRevC.61.054906} {\bibfield  {journal} {\bibinfo
  {journal} {Phys. Rev. C}\ }\textbf {\bibinfo {volume} {61}},\ \bibinfo
  {pages} {054906} (\bibinfo {year} {2000})}\BibitemShut {NoStop}%
\bibitem [{\citenamefont {Vitev}(2007)}]{Vitev:2007ve}%
  \BibitemOpen
  \bibfield  {author} {\bibinfo {author} {\bibfnamefont {I.}~\bibnamefont
  {Vitev}},\ }\bibfield  {title} {\bibinfo {title} {{Non-Abelian energy loss in
  cold nuclear matter}},\ }\href {https://doi.org/10.1103/PhysRevC.75.064906}
  {\bibfield  {journal} {\bibinfo  {journal} {Phys. Rev. C}\ }\textbf {\bibinfo
  {volume} {75}},\ \bibinfo {pages} {064906} (\bibinfo {year}
  {2007})}\BibitemShut {NoStop}%
\bibitem [{\citenamefont {Cronin}\ \emph {et~al.}(1975)\citenamefont {Cronin},
  \citenamefont {Frisch}, \citenamefont {Shochet}, \citenamefont {Boymond},
  \citenamefont {Mermod}, \citenamefont {Piroue},\ and\ \citenamefont
  {Sumner}}]{Cronin:1974zm}%
  \BibitemOpen
  \bibfield  {author} {\bibinfo {author} {\bibfnamefont {J.~W.}\ \bibnamefont
  {Cronin}}, \bibinfo {author} {\bibfnamefont {H.~J.}\ \bibnamefont {Frisch}},
  \bibinfo {author} {\bibfnamefont {M.~J.}\ \bibnamefont {Shochet}}, \bibinfo
  {author} {\bibfnamefont {J.~P.}\ \bibnamefont {Boymond}}, \bibinfo {author}
  {\bibfnamefont {R.}~\bibnamefont {Mermod}}, \bibinfo {author} {\bibfnamefont
  {P.~A.}\ \bibnamefont {Piroue}},\ and\ \bibinfo {author} {\bibfnamefont
  {R.~L.}\ \bibnamefont {Sumner}},\ }\bibfield  {title} {\bibinfo {title}
  {{Production of hadrons with large transverse momentum at 200, 300, and 400
  GeV}},\ }\href {https://doi.org/10.1103/PhysRevD.11.3105} {\bibfield
  {journal} {\bibinfo  {journal} {Phys. Rev. D}\ }\textbf {\bibinfo {volume}
  {11}},\ \bibinfo {pages} {3105} (\bibinfo {year} {1975})}\BibitemShut
  {NoStop}%
\bibitem [{\citenamefont {Adam}\ \emph {et~al.}({\natexlab{a}})\citenamefont
  {Adam} \emph {et~al.}}]{ALICE:2015kgk}%
  \BibitemOpen
  \bibfield  {author} {\bibinfo {author} {\bibfnamefont {J.}~\bibnamefont
  {Adam}} \emph {et~al.} (\bibinfo {collaboration} {ALICE Collaboration}),\
  }\bibfield  {title} {\bibinfo {title} {{Centrality dependence of inclusive
  $J_\psi$ production in $p$$-$Pb collisions at $\sqrt{s_{_{NN}}}=5.02$\,TeV}}}
  ({\natexlab{a}}),\ \bibinfo {note} {{J. High Energy Phys. {\bf 11 (2015)},
  127}}\BibitemShut {NoStop}%
\bibitem [{\citenamefont {Abelev}\ \emph {et~al.}({\natexlab{b}})\citenamefont
  {Abelev} \emph {et~al.}}]{ALICE:2013snh}%
  \BibitemOpen
  \bibfield  {author} {\bibinfo {author} {\bibfnamefont {B.~B.}\ \bibnamefont
  {Abelev}} \emph {et~al.} (\bibinfo {collaboration} {ALICE Collaboration}),\
  }\bibfield  {title} {\bibinfo {title} {{$J/\psi$ production and nuclear
  effects in $p$-Pb collisions at $\sqrt{S_{NN}}$ = 5.02 TeV}}}
  ({\natexlab{b}}),\ \bibinfo {note} {{J. High Energy Phys. {\bf 02 (2014)},
  073}}\BibitemShut {NoStop}%
\bibitem [{\citenamefont {Aad}\ \emph {et~al.}(2015)\citenamefont {Aad} \emph
  {et~al.}}]{Aad:2015ddl}%
  \BibitemOpen
  \bibfield  {author} {\bibinfo {author} {\bibfnamefont {G.}~\bibnamefont
  {Aad}} \emph {et~al.} (\bibinfo {collaboration} {ATLAS Collaboration}),\
  }\bibfield  {title} {\bibinfo {title} {{Measurement of differential $J/\psi$
  production cross sections and forward-backward ratios in $p$$+$Pb collisions
  with the ATLAS detector}},\ }\href
  {https://doi.org/10.1103/PhysRevC.92.034904} {\bibfield  {journal} {\bibinfo
  {journal} {Phys. Rev. C}\ }\textbf {\bibinfo {volume} {92}},\ \bibinfo
  {pages} {034904} (\bibinfo {year} {2015})}\BibitemShut {NoStop}%
\bibitem [{\citenamefont {Aaboud}\ \emph {et~al.}(2018)\citenamefont {Aaboud}
  \emph {et~al.}}]{Aaboud:2017cif}%
  \BibitemOpen
  \bibfield  {author} {\bibinfo {author} {\bibfnamefont {M.}~\bibnamefont
  {Aaboud}} \emph {et~al.} (\bibinfo {collaboration} {ATLAS Collaboration}),\
  }\bibfield  {title} {\bibinfo {title} {{Measurement of quarkonium production
  in proton-lead and proton-proton collisions at $5.02~\mathrm {TeV}$ with the
  ATLAS detector}},\ }\href {https://doi.org/10.1140/epjc/s10052-018-5624-4}
  {\bibfield  {journal} {\bibinfo  {journal} {Eur. Phys. J. C}\ }\textbf
  {\bibinfo {volume} {78}},\ \bibinfo {pages} {171} (\bibinfo {year}
  {2018})}\BibitemShut {NoStop}%
\bibitem [{\citenamefont {Adam}\ \emph {et~al.}({\natexlab{b}})\citenamefont
  {Adam} \emph {et~al.}}]{ALICE:2016sdt}%
  \BibitemOpen
  \bibfield  {author} {\bibinfo {author} {\bibfnamefont {J.}~\bibnamefont
  {Adam}} \emph {et~al.} (\bibinfo {collaboration} {ALICE Collaboration}),\
  }\bibfield  {title} {\bibinfo {title} {{Centrality dependence of
  ${\psi}\,(2S)$ suppression in $p$$-$Pb collisions at
  $\sqrt{s_{_{NN}}}=5.02$\,TeV}}} ({\natexlab{b}}),\ \bibinfo {note} {{J. High
  Energy Phys. {\bf 06 (2016)}, 050}}\BibitemShut {NoStop}%
\bibitem [{\citenamefont {Acharya}\ \emph {et~al.}()\citenamefont {Acharya}
  \emph {et~al.}}]{Acharya:2020rvc}%
  \BibitemOpen
  \bibfield  {author} {\bibinfo {author} {\bibfnamefont {S.}~\bibnamefont
  {Acharya}} \emph {et~al.} (\bibinfo {collaboration} {ALICE Collaboration}),\
  }\bibfield  {title} {\bibinfo {title} {{Centrality dependence of J/$\psi$ and
  $\psi$(2S) production and nuclear modification in p-Pb collisions at
  $\sqrt{s_{\rm NN}} =$ 8.16 TeV}},\ }\bibinfo {note} {{J. High Energy Phys.
  {\bf 02 (2021)}, 002}}\BibitemShut {NoStop}%
\bibitem [{\citenamefont {Aaij}\ \emph {et~al.}({\natexlab{b}})\citenamefont
  {Aaij} \emph {et~al.}}]{LHCb:2013gmv}%
  \BibitemOpen
  \bibfield  {author} {\bibinfo {author} {\bibfnamefont {R.}~\bibnamefont
  {Aaij}} \emph {et~al.} (\bibinfo {collaboration} {LHCb Collaboration}),\
  }\bibfield  {title} {\bibinfo {title} {{Study of $J/\psi$ production and cold
  nuclear matter effects in $pPb$ collisions at $\sqrt{s_{NN}} = 5$ TeV}}}
  ({\natexlab{b}}),\ \bibinfo {note} {{J. High Energy Phys. {\bf 02 (2014)},
  072}}\BibitemShut {NoStop}%
\bibitem [{\citenamefont {Sirunyan}\ \emph {et~al.}(2017)\citenamefont
  {Sirunyan} \emph {et~al.}}]{Sirunyan:2017mzd}%
  \BibitemOpen
  \bibfield  {author} {\bibinfo {author} {\bibfnamefont {A.~M.}\ \bibnamefont
  {Sirunyan}} \emph {et~al.} (\bibinfo {collaboration} {CMS Collaboration}),\
  }\bibfield  {title} {\bibinfo {title} {{Measurement of prompt and nonprompt
  $J/{\psi}$ production in $pp$ and $ p\mathrm {Pb}$ collisions at
  $\sqrt{s_{\mathrm {NN}}} =5.02\,\text {TeV} $}},\ }\href
  {https://doi.org/10.1140/epjc/s10052-017-4828-3} {\bibfield  {journal}
  {\bibinfo  {journal} {Eur. Phys. J. C}\ }\textbf {\bibinfo {volume} {77}},\
  \bibinfo {pages} {269} (\bibinfo {year} {2017})}\BibitemShut {NoStop}%
\bibitem [{\citenamefont {Sirunyan}\ \emph {et~al.}(2019)\citenamefont
  {Sirunyan} \emph {et~al.}}]{Sirunyan:2018pse}%
  \BibitemOpen
  \bibfield  {author} {\bibinfo {author} {\bibfnamefont {A.~M.}\ \bibnamefont
  {Sirunyan}} \emph {et~al.} (\bibinfo {collaboration} {CMS Collaboration}),\
  }\bibfield  {title} {\bibinfo {title} {{Measurement of prompt $\psi(2S)$
  production cross sections in proton-lead and proton-proton collisions at
  $\sqrt{s_{_\mathrm{NN}}}=$ 5.02 TeV}},\ }\href
  {https://doi.org/10.1016/j.physletb.2019.01.058} {\bibfield  {journal}
  {\bibinfo  {journal} {Phys. Lett. B}\ }\textbf {\bibinfo {volume} {790}},\
  \bibinfo {pages} {509} (\bibinfo {year} {2019})}\BibitemShut {NoStop}%
\bibitem [{\citenamefont {Adare}\ \emph {et~al.}(2012)\citenamefont {Adare}
  \emph {et~al.}}]{PHENIX:2011gyb}%
  \BibitemOpen
  \bibfield  {author} {\bibinfo {author} {\bibfnamefont {A.}~\bibnamefont
  {Adare}} \emph {et~al.} (\bibinfo {collaboration} {PHENIX Collaboration}),\
  }\bibfield  {title} {\bibinfo {title} {{Ground and excited charmonium state
  production in $p+p$ collisions at $\sqrt{s}=200$ GeV}},\ }\href
  {https://doi.org/10.1103/PhysRevD.85.092004} {\bibfield  {journal} {\bibinfo
  {journal} {Phys. Rev. D}\ }\textbf {\bibinfo {volume} {85}},\ \bibinfo
  {pages} {092004} (\bibinfo {year} {2012})}\BibitemShut {NoStop}%
\bibitem [{\citenamefont {Adare}\ \emph
  {et~al.}(2011{\natexlab{a}})\citenamefont {Adare} \emph
  {et~al.}}]{Adare:2010fn}%
  \BibitemOpen
  \bibfield  {author} {\bibinfo {author} {\bibfnamefont {A.}~\bibnamefont
  {Adare}} \emph {et~al.} (\bibinfo {collaboration} {PHENIX Collaboration}),\
  }\bibfield  {title} {\bibinfo {title} {{Cold Nuclear Matter Effects on
  $J/\psi$ Yields as a Function of Rapidity and Nuclear Geometry in
  Deuteron-Gold Collisions at $\sqrt{s_{NN}}=200$ GeV}},\ }\href
  {https://doi.org/10.1103/PhysRevLett.107.142301} {\bibfield  {journal}
  {\bibinfo  {journal} {Phys. Rev. Lett.}\ }\textbf {\bibinfo {volume} {107}},\
  \bibinfo {pages} {142301} (\bibinfo {year} {2011}{\natexlab{a}})}\BibitemShut
  {NoStop}%
\bibitem [{\citenamefont {Acharya}\ \emph
  {et~al.}(2020{\natexlab{a}})\citenamefont {Acharya} \emph
  {et~al.}}]{Acharya:2019zjt}%
  \BibitemOpen
  \bibfield  {author} {\bibinfo {author} {\bibfnamefont {U.}~\bibnamefont
  {Acharya}} \emph {et~al.} (\bibinfo {collaboration} {PHENIX Collaboration}),\
  }\bibfield  {title} {\bibinfo {title} {{Measurement of $J/\psi$ at forward
  and backward rapidity in $p+p$, $p+$Al, $p+$Au, and $^3$He$+$Au collisions at
  $\sqrt{s_{_{NN}}}=200~{\rm GeV}$}},\ }\href
  {https://doi.org/10.1103/PhysRevC.102.014902} {\bibfield  {journal} {\bibinfo
   {journal} {Phys. Rev. C}\ }\textbf {\bibinfo {volume} {102}},\ \bibinfo
  {pages} {014902} (\bibinfo {year} {2020}{\natexlab{a}})}\BibitemShut
  {NoStop}%
\bibitem [{\citenamefont {Adamczyk}\ \emph
  {et~al.}(2016{\natexlab{b}})\citenamefont {Adamczyk} \emph
  {et~al.}}]{Adamczyk:2016dhc}%
  \BibitemOpen
  \bibfield  {author} {\bibinfo {author} {\bibfnamefont {L.}~\bibnamefont
  {Adamczyk}} \emph {et~al.} (\bibinfo {collaboration} {STAR Collaboration}),\
  }\bibfield  {title} {\bibinfo {title} {{$J/\psi$ production at low transverse
  momentum in $p$+$p$ and $d$+Au collisions at $\sqrt{s_{NN}}$ = 200 GeV}},\
  }\href {https://doi.org/10.1103/PhysRevC.93.064904} {\bibfield  {journal}
  {\bibinfo  {journal} {Phys. Rev. C}\ }\textbf {\bibinfo {volume} {93}},\
  \bibinfo {pages} {064904} (\bibinfo {year} {2016}{\natexlab{b}})}\BibitemShut
  {NoStop}%
\bibitem [{\citenamefont {Adare}\ \emph {et~al.}(2017)\citenamefont {Adare}
  \emph {et~al.}}]{PHENIX:2016vmz}%
  \BibitemOpen
  \bibfield  {author} {\bibinfo {author} {\bibfnamefont {A.}~\bibnamefont
  {Adare}} \emph {et~al.} (\bibinfo {collaboration} {PHENIX Collaboration}),\
  }\bibfield  {title} {\bibinfo {title} {{Measurement of the relative yields of
  $\psi(2S)$ to $\psi(1S)$ mesons produced at forward and backward rapidity in
  $p+p$, $p+$Al, $p+$Au, and $^{3}$He$+$Au collisions at $\sqrt{s_{{NN}}}=200$
  GeV}},\ }\href {https://doi.org/10.1103/PhysRevC.95.034904} {\bibfield
  {journal} {\bibinfo  {journal} {Phys. Rev. C}\ }\textbf {\bibinfo {volume}
  {95}},\ \bibinfo {pages} {034904} (\bibinfo {year} {2017})}\BibitemShut
  {NoStop}%
\bibitem [{\citenamefont {Akikawa}\ \emph {et~al.}(2003)\citenamefont {Akikawa}
  \emph {et~al.}}]{Akikawa:2003zs}%
  \BibitemOpen
  \bibfield  {author} {\bibinfo {author} {\bibfnamefont {H.}~\bibnamefont
  {Akikawa}} \emph {et~al.} (\bibinfo {collaboration} {PHENIX Collaboration}),\
  }\bibfield  {title} {\bibinfo {title} {{PHENIX muon arms}},\ }\href
  {https://doi.org/10.1016/S0168-9002(02)01955-1} {\bibfield  {journal}
  {\bibinfo  {journal} {Nucl. Instrum. Methods Phys. Res., Sec. A}\ }\textbf
  {\bibinfo {volume} {499}},\ \bibinfo {pages} {537} (\bibinfo {year}
  {2003})}\BibitemShut {NoStop}%
\bibitem [{\citenamefont {Adachi}\ \emph {et~al.}(2013)\citenamefont {Adachi}
  \emph {et~al.}}]{Adachi:2013qha}%
  \BibitemOpen
  \bibfield  {author} {\bibinfo {author} {\bibfnamefont {S.}~\bibnamefont
  {Adachi}} \emph {et~al.},\ }\bibfield  {title} {\bibinfo {title} {{Trigger
  electronics upgrade of PHENIX muon tracker}},\ }\href
  {https://doi.org/10.1016/j.nima.2012.11.088} {\bibfield  {journal} {\bibinfo
  {journal} {Nucl. Instrum. Methods Phys. Res., Sec. A}\ }\textbf {\bibinfo
  {volume} {703}},\ \bibinfo {pages} {114} (\bibinfo {year}
  {2013})}\BibitemShut {NoStop}%
\bibitem [{\citenamefont {Aidala}\ \emph {et~al.}(2014)\citenamefont {Aidala}
  \emph {et~al.}}]{Aidala:2013vna}%
  \BibitemOpen
  \bibfield  {author} {\bibinfo {author} {\bibfnamefont {C.}~\bibnamefont
  {Aidala}} \emph {et~al.},\ }\bibfield  {title} {\bibinfo {title} {{The PHENIX
  Forward Silicon Vertex Detector}},\ }\href
  {https://doi.org/10.1016/j.nima.2014.04.017} {\bibfield  {journal} {\bibinfo
  {journal} {Nucl. Instrum. Methods Phys. Res., Sec. A}\ }\textbf {\bibinfo
  {volume} {755}},\ \bibinfo {pages} {44} (\bibinfo {year} {2014})}\BibitemShut
  {NoStop}%
\bibitem [{\citenamefont {Allen}\ \emph {et~al.}(2003)\citenamefont {Allen}
  \emph {et~al.}}]{Allen:2003zt}%
  \BibitemOpen
  \bibfield  {author} {\bibinfo {author} {\bibfnamefont {M.}~\bibnamefont
  {Allen}} \emph {et~al.} (\bibinfo {collaboration} {PHENIX Collaboration}),\
  }\bibfield  {title} {\bibinfo {title} {{PHENIX inner detectors}},\ }\href
  {https://doi.org/10.1016/S0168-9002(02)01956-3} {\bibfield  {journal}
  {\bibinfo  {journal} {Nucl. Instrum. Methods Phys. Res., Sec. A}\ }\textbf
  {\bibinfo {volume} {499}},\ \bibinfo {pages} {549} (\bibinfo {year}
  {2003})}\BibitemShut {NoStop}%
\bibitem [{\citenamefont {Adare}\ \emph {et~al.}(2014)\citenamefont {Adare}
  \emph {et~al.}}]{PHENIX:2013mpn}%
  \BibitemOpen
  \bibfield  {author} {\bibinfo {author} {\bibfnamefont {A.}~\bibnamefont
  {Adare}} \emph {et~al.} (\bibinfo {collaboration} {PHENIX Collaboration}),\
  }\bibfield  {title} {\bibinfo {title} {{Centrality categorization for
  $R_{p(d)+A}$ in high-energy collisions}},\ }\href
  {https://doi.org/10.1103/PhysRevC.90.034902} {\bibfield  {journal} {\bibinfo
  {journal} {Phys. Rev. C}\ }\textbf {\bibinfo {volume} {90}},\ \bibinfo
  {pages} {034902} (\bibinfo {year} {2014})}\BibitemShut {NoStop}%
\bibitem [{\citenamefont {Gaiser}(1982)}]{Gaiser:1982yw}%
  \BibitemOpen
  \bibfield  {author} {\bibinfo {author} {\bibfnamefont {J.~E.}\ \bibnamefont
  {Gaiser}},\ }\emph {\bibinfo {title} {{Charmonium Spectroscopy From Radiative
  Decays of the $J/\psi$ and $\psi^\prime$}}},\ \href
  {http://www-public.slac.stanford.edu/sciDoc/docMeta.aspx?slacPubNumber=slac-r-255.html}
  {Ph.D. thesis},\ \bibinfo  {school} {SLAC} (\bibinfo {year}
  {1982})\BibitemShut {NoStop}%
\bibitem [{\citenamefont {Aidala}\ \emph
  {et~al.}(2019{\natexlab{b}})\citenamefont {Aidala} \emph
  {et~al.}}]{Aidala:2018ajl}%
  \BibitemOpen
  \bibfield  {author} {\bibinfo {author} {\bibfnamefont {C.}~\bibnamefont
  {Aidala}} \emph {et~al.} (\bibinfo {collaboration} {PHENIX Collaboration}),\
  }\bibfield  {title} {\bibinfo {title} {{Measurements of $\mu\mu$ pairs from
  open heavy flavor and Drell-Yan in $p$$+$$p$ collisions at $\sqrt{s}=200$
  GeV}},\ }\href {https://doi.org/10.1103/PhysRevD.99.072003} {\bibfield
  {journal} {\bibinfo  {journal} {Phys. Rev. D}\ }\textbf {\bibinfo {volume}
  {99}},\ \bibinfo {pages} {072003} (\bibinfo {year}
  {2019}{\natexlab{b}})}\BibitemShut {NoStop}%
\bibitem [{\citenamefont {Adare}\ \emph
  {et~al.}(2011{\natexlab{b}})\citenamefont {Adare} \emph
  {et~al.}}]{Adare:2010de}%
  \BibitemOpen
  \bibfield  {author} {\bibinfo {author} {\bibfnamefont {A.}~\bibnamefont
  {Adare}} \emph {et~al.} (\bibinfo {collaboration} {PHENIX Collaboration}),\
  }\bibfield  {title} {\bibinfo {title} {{Heavy Quark Production in $p$$+$$p$
  and Energy Loss and Flow of Heavy Quarks in Au$+$Au Collisions at
  $\sqrt{s_{NN}}=200$ GeV}},\ }\href
  {https://doi.org/10.1103/PhysRevC.84.044905} {\bibfield  {journal} {\bibinfo
  {journal} {Phys. Rev. C}\ }\textbf {\bibinfo {volume} {84}},\ \bibinfo
  {pages} {044905} (\bibinfo {year} {2011}{\natexlab{b}})}\BibitemShut
  {NoStop}%
\bibitem [{\citenamefont {Choudhury}\ \emph {et~al.}(2007)\citenamefont
  {Choudhury} \emph {et~al.}}]{osti:912839}%
  \BibitemOpen
  \bibfield  {author} {\bibinfo {author} {\bibfnamefont {R.}~\bibnamefont
  {Choudhury}} \emph {et~al.},\ }\bibfield  {title} {\bibinfo {title}
  {Technical design report of the forward silicon vertex tracker}} (\bibinfo
  {year} {2007}),\ \bibinfo {note}
  {https://www.osti.gov/biblio/912839}\BibitemShut {NoStop}%
\bibitem [{\citenamefont {Tanabashi}\ \emph {et~al.}(2018)\citenamefont
  {Tanabashi} \emph {et~al.}}]{PDG:2018}%
  \BibitemOpen
  \bibfield  {author} {\bibinfo {author} {\bibfnamefont {M.}~\bibnamefont
  {Tanabashi}} \emph {et~al.} (\bibinfo {collaboration} {Particle Data
  Group}),\ }\bibfield  {title} {\bibinfo {title} {{Review of Particle
  Physics}},\ }\href {https://doi.org/10.1103/PhysRevD.98.030001} {\bibfield
  {journal} {\bibinfo  {journal} {Phys. Rev. D}\ }\textbf {\bibinfo {volume}
  {98}},\ \bibinfo {pages} {030001} (\bibinfo {year} {2018})}\BibitemShut
  {NoStop}%
\bibitem [{\citenamefont {Leung}(2019)}]{Leung:2019vbb}%
  \BibitemOpen
  \bibfield  {author} {\bibinfo {author} {\bibfnamefont {Y.~H.}\ \bibnamefont
  {Leung}} (\bibinfo {collaboration} {PHENIX Collaboration}),\ }\bibfield
  {title} {\bibinfo {title} {{Measurements of charm, bottom, and Drell-Yan via
  dimuons in $p$$+$$p$ and $p$$+$Au collisions at $\sqrt {s_{NN}}=200$~GeV
  with PHENIX at RHIC}},\ }\href
  {https://doi.org/10.1016/j.nuclphysa.2018.09.043} {\bibfield  {journal}
  {\bibinfo  {journal} {Nucl. Phys. A}\ }\textbf {\bibinfo {volume} {982}},\
  \bibinfo {pages} {695} (\bibinfo {year} {2019})}\BibitemShut {NoStop}%
\bibitem [{\citenamefont {Sj$\mathrm{\ddot{o}}$strand}\ \emph
  {et~al.}(2008)\citenamefont {Sj$\mathrm{\ddot{o}}$strand}, \citenamefont
  {Mrenna},\ and\ \citenamefont {Skands}}]{PYTHIA:81}%
  \BibitemOpen
  \bibfield  {author} {\bibinfo {author} {\bibfnamefont {T.}~\bibnamefont
  {Sj$\mathrm{\ddot{o}}$strand}}, \bibinfo {author} {\bibfnamefont
  {S.}~\bibnamefont {Mrenna}},\ and\ \bibinfo {author} {\bibfnamefont {P.~Z.}\
  \bibnamefont {Skands}},\ }\bibfield  {title} {\bibinfo {title} {{A Brief
  Introduction to PYTHIA 8.1}},\ }\href
  {https://doi.org/10.1016/j.cpc.2008.01.036} {\bibfield  {journal} {\bibinfo
  {journal} {Comput. Phys. Commun.}\ }\textbf {\bibinfo {volume} {178}},\
  \bibinfo {pages} {852} (\bibinfo {year} {2008})}\BibitemShut {NoStop}%
\bibitem [{\citenamefont {Agostinelli}\ \emph {et~al.}(2003)\citenamefont
  {Agostinelli} \emph {et~al.}}]{Geant:4}%
  \BibitemOpen
  \bibfield  {author} {\bibinfo {author} {\bibfnamefont {S.}~\bibnamefont
  {Agostinelli}} \emph {et~al.} (\bibinfo {collaboration} {GEANT4
  Collaboration}),\ }\bibfield  {title} {\bibinfo {title} {{GEANT4: A
  Simulation toolkit}},\ }\href {https://doi.org/10.1016/S0168-9002(03)01368-8}
  {\bibfield  {journal} {\bibinfo  {journal} {Nucl. Instrum. Methods Phys.
  Res., Sec. A}\ }\textbf {\bibinfo {volume} {506}},\ \bibinfo {pages} {250}
  (\bibinfo {year} {2003})}\BibitemShut {NoStop}%
\bibitem [{\citenamefont {Drees}\ \emph {et~al.}(2003)\citenamefont {Drees},
  \citenamefont {Fox}, \citenamefont {Xu},\ and\ \citenamefont
  {Huang}}]{Drees:2003zza}%
  \BibitemOpen
  \bibfield  {author} {\bibinfo {author} {\bibfnamefont {A.}~\bibnamefont
  {Drees}}, \bibinfo {author} {\bibfnamefont {B.}~\bibnamefont {Fox}}, \bibinfo
  {author} {\bibfnamefont {Z.}~\bibnamefont {Xu}},\ and\ \bibinfo {author}
  {\bibfnamefont {H.}~\bibnamefont {Huang}},\ }\bibfield  {title} {\bibinfo
  {title} {{Results from Vernier Scans at RHIC during the $pp$ Run
  2001-2002}},\ }\href@noop {} {\bibfield  {journal} {\bibinfo  {journal}
  {Conf. Proc.}\ }\textbf {\bibinfo {volume} {C030512}},\ \bibinfo {pages}
  {1688} (\bibinfo {year} {2003})}\BibitemShut {NoStop}%
\bibitem [{\citenamefont {Aidala}\ \emph {et~al.}(2017)\citenamefont {Aidala}
  \emph {et~al.}}]{PHENIX:2017caf}%
  \BibitemOpen
  \bibfield  {author} {\bibinfo {author} {\bibfnamefont {C.}~\bibnamefont
  {Aidala}} \emph {et~al.} (\bibinfo {collaboration} {PHENIX Collaboration}),\
  }\bibfield  {title} {\bibinfo {title} {{$B$-meson production at forward and
  backward rapidity in $p$$+$$p$ and Cu$+$Au collisions at
  $\sqrt{s_{_{NN}}}=200$ GeV}},\ }\href
  {https://doi.org/10.1103/PhysRevC.96.064901} {\bibfield  {journal} {\bibinfo
  {journal} {Phys. Rev. C}\ }\textbf {\bibinfo {volume} {96}},\ \bibinfo
  {pages} {064901} (\bibinfo {year} {2017})}\BibitemShut {NoStop}%
\bibitem [{\citenamefont {Adler}\ \emph {et~al.}(2003)\citenamefont {Adler}
  \emph {et~al.}}]{Adler:2003pb}%
  \BibitemOpen
  \bibfield  {author} {\bibinfo {author} {\bibfnamefont {S.~S.}\ \bibnamefont
  {Adler}} \emph {et~al.} (\bibinfo {collaboration} {PHENIX Collaboration}),\
  }\bibfield  {title} {\bibinfo {title} {{Mid-rapidity neutral pion production
  in proton proton collisions at $\sqrt{s}=200~\mathrm{GeV}$}},\ }\href
  {https://doi.org/10.1103/PhysRevLett.91.241803} {\bibfield  {journal}
  {\bibinfo  {journal} {Phys. Rev. Lett.}\ }\textbf {\bibinfo {volume} {91}},\
  \bibinfo {pages} {241803} (\bibinfo {year} {2003})}\BibitemShut {NoStop}%
\bibitem [{\citenamefont {Acharya}\ \emph
  {et~al.}(2020{\natexlab{b}})\citenamefont {Acharya} \emph
  {et~al.}}]{PHENIX:2020dqu}%
  \BibitemOpen
  \bibfield  {author} {\bibinfo {author} {\bibfnamefont {U.}~\bibnamefont
  {Acharya}} \emph {et~al.} (\bibinfo {collaboration} {PHENIX Collaboration}),\
  }\bibfield  {title} {\bibinfo {title} {{Polarization and cross section of
  midrapidity J/$\psi$ production in $p+p$ collisions at $\sqrt {s}$ = 510
  GeV}},\ }\href {https://doi.org/10.1103/PhysRevD.102.072008} {\bibfield
  {journal} {\bibinfo  {journal} {Phys. Rev. D}\ }\textbf {\bibinfo {volume}
  {102}},\ \bibinfo {pages} {072008} (\bibinfo {year}
  {2020}{\natexlab{b}})}\BibitemShut {NoStop}%
\bibitem [{\citenamefont {Abelev}\ \emph {et~al.}(2012)\citenamefont {Abelev}
  \emph {et~al.}}]{ALICE:2011gej}%
  \BibitemOpen
  \bibfield  {author} {\bibinfo {author} {\bibfnamefont {B.}~\bibnamefont
  {Abelev}} \emph {et~al.} (\bibinfo {collaboration} {ALICE Collaboration}),\
  }\bibfield  {title} {\bibinfo {title} {{$J/\psi$ polarization in $pp$
  collisions at $\sqrt{s}=7$ TeV}},\ }\href
  {https://doi.org/10.1103/PhysRevLett.108.082001} {\bibfield  {journal}
  {\bibinfo  {journal} {Phys. Rev. Lett.}\ }\textbf {\bibinfo {volume} {108}},\
  \bibinfo {pages} {082001} (\bibinfo {year} {2012})}\BibitemShut {NoStop}%
\bibitem [{\citenamefont {Aaij}\ \emph {et~al.}(2013)\citenamefont {Aaij} \emph
  {et~al.}}]{LHCb:2013izl}%
  \BibitemOpen
  \bibfield  {author} {\bibinfo {author} {\bibfnamefont {R.}~\bibnamefont
  {Aaij}} \emph {et~al.} (\bibinfo {collaboration} {LHCb Collaboration}),\
  }\bibfield  {title} {\bibinfo {title} {{Measurement of $J/\psi$ polarization
  in $pp$ collisions at $\sqrt{s}=7$ TeV}},\ }\href
  {https://doi.org/10.1140/epjc/s10052-013-2631-3} {\bibfield  {journal}
  {\bibinfo  {journal} {Eur. Phys. J. C}\ }\textbf {\bibinfo {volume} {73}},\
  \bibinfo {pages} {2631} (\bibinfo {year} {2013})}\BibitemShut {NoStop}%
\bibitem [{\citenamefont {Chatrchyan}\ \emph {et~al.}(2013)\citenamefont
  {Chatrchyan} \emph {et~al.}}]{CMS:2013gbz}%
  \BibitemOpen
  \bibfield  {author} {\bibinfo {author} {\bibfnamefont {S.}~\bibnamefont
  {Chatrchyan}} \emph {et~al.} (\bibinfo {collaboration} {CMS Collaboration}),\
  }\bibfield  {title} {\bibinfo {title} {{Measurement of the Prompt $J/\psi$
  and $\psi$(2S) Polarizations in $pp$ Collisions at $\sqrt{s}$ = 7 TeV}},\
  }\href {https://doi.org/10.1016/j.physletb.2013.10.055} {\bibfield  {journal}
  {\bibinfo  {journal} {Phys. Lett. B}\ }\textbf {\bibinfo {volume} {727}},\
  \bibinfo {pages} {381} (\bibinfo {year} {2013})}\BibitemShut {NoStop}%
\bibitem [{\citenamefont {Du}\ and\ \citenamefont {Rapp}()}]{Du:2018wsj}%
  \BibitemOpen
  \bibfield  {author} {\bibinfo {author} {\bibfnamefont {X.}~\bibnamefont
  {Du}}\ and\ \bibinfo {author} {\bibfnamefont {R.}~\bibnamefont {Rapp}},\
  }\bibfield  {title} {\bibinfo {title} {{In-Medium Charmonium Production in
  Proton-Nucleus Collisions}},\ }\bibinfo {note} {{J. High Energy Phys. {\bf 03
  (2019)}, 015}}\BibitemShut {NoStop}%
\bibitem [{\citenamefont {Eskola}\ \emph {et~al.}()\citenamefont {Eskola},
  \citenamefont {Paukkunen},\ and\ \citenamefont {Salgado}}]{Eskola:2009uj}%
  \BibitemOpen
  \bibfield  {author} {\bibinfo {author} {\bibfnamefont {K.~J.}\ \bibnamefont
  {Eskola}}, \bibinfo {author} {\bibfnamefont {H.}~\bibnamefont {Paukkunen}},\
  and\ \bibinfo {author} {\bibfnamefont {C.~A.}\ \bibnamefont {Salgado}},\
  }\bibfield  {title} {\bibinfo {title} {{EPS09: A New Generation of NLO and LO
  Nuclear Parton Distribution Functions}},\ }\bibinfo {note} {{J. High Energy
  Phys. {\bf 04 (2009)}, 065}}\BibitemShut {NoStop}%
\bibitem [{\citenamefont {Vogt}(2010)}]{Vogt:2010aa}%
  \BibitemOpen
  \bibfield  {author} {\bibinfo {author} {\bibfnamefont {R.}~\bibnamefont
  {Vogt}},\ }\bibfield  {title} {\bibinfo {title} {{Cold Nuclear Matter Effects
  on $J/\psi$ and $\Upsilon$ Production at the LHC}},\ }\href
  {https://doi.org/10.1103/PhysRevC.81.044903} {\bibfield  {journal} {\bibinfo
  {journal} {Phys. Rev. C}\ }\textbf {\bibinfo {volume} {81}},\ \bibinfo
  {pages} {044903} (\bibinfo {year} {2010})}\BibitemShut {NoStop}%
\bibitem [{\citenamefont {Adare}\ \emph {et~al.}(2018)\citenamefont {Adare}
  \emph {et~al.}}]{PHENIX:2018hho}%
  \BibitemOpen
  \bibfield  {author} {\bibinfo {author} {\bibfnamefont {A.}~\bibnamefont
  {Adare}} \emph {et~al.} (\bibinfo {collaboration} {PHENIX Collaboration}),\
  }\bibfield  {title} {\bibinfo {title} {{Pseudorapidity Dependence of Particle
  Production and Elliptic Flow in Asymmetric Nuclear Collisions of $p+$Al,
  $p+$Au, $d+$Au, and $^{3}$He$+$Au at $\sqrt{s_{_{NN}}}=200$ GeV}},\ }\href
  {https://doi.org/10.1103/PhysRevLett.121.222301} {\bibfield  {journal}
  {\bibinfo  {journal} {Phys. Rev. Lett.}\ }\textbf {\bibinfo {volume} {121}},\
  \bibinfo {pages} {222301} (\bibinfo {year} {2018})}\BibitemShut {NoStop}%
\bibitem [{\citenamefont {Kusina}\ \emph {et~al.}(2018)\citenamefont {Kusina},
  \citenamefont {Lansberg}, \citenamefont {Schienbein},\ and\ \citenamefont
  {Shao}}]{Kusina:2017gkz}%
  \BibitemOpen
  \bibfield  {author} {\bibinfo {author} {\bibfnamefont {A.}~\bibnamefont
  {Kusina}}, \bibinfo {author} {\bibfnamefont {J.-P.}\ \bibnamefont
  {Lansberg}}, \bibinfo {author} {\bibfnamefont {I.}~\bibnamefont
  {Schienbein}},\ and\ \bibinfo {author} {\bibfnamefont {H.-S.}\ \bibnamefont
  {Shao}},\ }\bibfield  {title} {\bibinfo {title} {{Gluon Shadowing in
  Heavy-Flavor Production at the LHC}},\ }\href
  {https://doi.org/10.1103/PhysRevLett.121.052004} {\bibfield  {journal}
  {\bibinfo  {journal} {Phys. Rev. Lett.}\ }\textbf {\bibinfo {volume} {121}},\
  \bibinfo {pages} {052004} (\bibinfo {year} {2018})}\BibitemShut {NoStop}%
\bibitem [{\citenamefont {Shao}(2013)}]{Shao:2012iz}%
  \BibitemOpen
  \bibfield  {author} {\bibinfo {author} {\bibfnamefont {H.-S.}\ \bibnamefont
  {Shao}},\ }\bibfield  {title} {\bibinfo {title} {{HELAC-Onia: An automatic
  matrix element generator for heavy quarkonium physics}},\ }\href
  {https://doi.org/10.1016/j.cpc.2013.05.023} {\bibfield  {journal} {\bibinfo
  {journal} {Comput. Phys. Commun.}\ }\textbf {\bibinfo {volume} {184}},\
  \bibinfo {pages} {2562} (\bibinfo {year} {2013})}\BibitemShut {NoStop}%
\bibitem [{\citenamefont {Shao}(2016)}]{Shao:2015vga}%
  \BibitemOpen
  \bibfield  {author} {\bibinfo {author} {\bibfnamefont {H.-S.}\ \bibnamefont
  {Shao}},\ }\bibfield  {title} {\bibinfo {title} {{HELAC-Onia 2.0: an upgraded
  matrix-element and event generator for heavy quarkonium physics}},\ }\href
  {https://doi.org/10.1016/j.cpc.2015.09.011} {\bibfield  {journal} {\bibinfo
  {journal} {Comput. Phys. Commun.}\ }\textbf {\bibinfo {volume} {198}},\
  \bibinfo {pages} {238} (\bibinfo {year} {2016})}\BibitemShut {NoStop}%
\bibitem [{\citenamefont {Lansberg}\ and\ \citenamefont
  {Shao}(2017)}]{Lansberg:2016deg}%
  \BibitemOpen
  \bibfield  {author} {\bibinfo {author} {\bibfnamefont {J.-P.}\ \bibnamefont
  {Lansberg}}\ and\ \bibinfo {author} {\bibfnamefont {H.-S.}\ \bibnamefont
  {Shao}},\ }\bibfield  {title} {\bibinfo {title} {{Towards an automated tool
  to evaluate the impact of the nuclear modification of the gluon density on
  quarkonium, D and B meson production in proton-nucleus collisions}},\ }\href
  {https://doi.org/10.1140/epjc/s10052-016-4575-x} {\bibfield  {journal}
  {\bibinfo  {journal} {Eur. Phys. J. C}\ }\textbf {\bibinfo {volume} {77}},\
  \bibinfo {pages} {1} (\bibinfo {year} {2017})}\BibitemShut {NoStop}%
\bibitem [{\citenamefont {Kusina}\ \emph {et~al.}(2021)\citenamefont {Kusina},
  \citenamefont {Lansberg}, \citenamefont {Schienbein},\ and\ \citenamefont
  {Shao}}]{Kusina:2020dki}%
  \BibitemOpen
  \bibfield  {author} {\bibinfo {author} {\bibfnamefont {A.}~\bibnamefont
  {Kusina}}, \bibinfo {author} {\bibfnamefont {J.-P.}\ \bibnamefont
  {Lansberg}}, \bibinfo {author} {\bibfnamefont {I.}~\bibnamefont
  {Schienbein}},\ and\ \bibinfo {author} {\bibfnamefont {H.-S.}\ \bibnamefont
  {Shao}},\ }\bibfield  {title} {\bibinfo {title} {{Reweighted nuclear PDFs
  using heavy-flavor production data at the LHC}},\ }\href
  {https://doi.org/10.1103/PhysRevD.104.014010} {\bibfield  {journal} {\bibinfo
   {journal} {Phys. Rev. D}\ }\textbf {\bibinfo {volume} {104}},\ \bibinfo
  {pages} {014010} (\bibinfo {year} {2021})}\BibitemShut {NoStop}%
\bibitem [{\citenamefont {Loizides}\ \emph {et~al.}(2018)\citenamefont
  {Loizides}, \citenamefont {Kamin},\ and\ \citenamefont
  {d'Enterria}}]{Loizides:2017ack}%
  \BibitemOpen
  \bibfield  {author} {\bibinfo {author} {\bibfnamefont {C.}~\bibnamefont
  {Loizides}}, \bibinfo {author} {\bibfnamefont {J.}~\bibnamefont {Kamin}},\
  and\ \bibinfo {author} {\bibfnamefont {D.}~\bibnamefont {d'Enterria}},\
  }\bibfield  {title} {\bibinfo {title} {{Improved Monte Carlo Glauber
  predictions at present and future nuclear colliders}},\ }\href
  {https://doi.org/10.1103/PhysRevC.97.054910} {\bibfield  {journal} {\bibinfo
  {journal} {Phys. Rev. C}\ }\textbf {\bibinfo {volume} {97}},\ \bibinfo
  {pages} {054910} (\bibinfo {year} {2018})},\ \bibinfo {note} {{\bf 99},
  019901(E) (2019)}\BibitemShut {NoStop}%
\bibitem [{\citenamefont {Shao}(2020)}]{Shao:2020acd}%
  \BibitemOpen
  \bibfield  {author} {\bibinfo {author} {\bibfnamefont {H.-S.}\ \bibnamefont
  {Shao}},\ }\bibfield  {title} {\bibinfo {title} {{Probing impact-parameter
  dependent nuclear parton densities from double parton scatterings in
  heavy-ion collisions}},\ }\href {https://doi.org/10.1103/PhysRevD.101.054036}
  {\bibfield  {journal} {\bibinfo  {journal} {Phys. Rev. D}\ }\textbf {\bibinfo
  {volume} {101}},\ \bibinfo {pages} {054036} (\bibinfo {year}
  {2020})}\BibitemShut {NoStop}%
\bibitem [{\citenamefont {Zhao}\ and\ \citenamefont
  {Rapp}(2010)}]{Zhao:2010nk}%
  \BibitemOpen
  \bibfield  {author} {\bibinfo {author} {\bibfnamefont {X.}~\bibnamefont
  {Zhao}}\ and\ \bibinfo {author} {\bibfnamefont {R.}~\bibnamefont {Rapp}},\
  }\bibfield  {title} {\bibinfo {title} {{Charmonium in Medium: From
  Correlators to Experiment}},\ }\href
  {https://doi.org/10.1103/PhysRevC.82.064905} {\bibfield  {journal} {\bibinfo
  {journal} {Phys. Rev. C}\ }\textbf {\bibinfo {volume} {82}},\ \bibinfo
  {pages} {064905} (\bibinfo {year} {2010})}\BibitemShut {NoStop}%
\bibitem [{\citenamefont {Abelev}\ \emph {et~al.}(2014)\citenamefont {Abelev}
  \emph {et~al.}}]{ALICE:2014uja}%
  \BibitemOpen
  \bibfield  {author} {\bibinfo {author} {\bibfnamefont {B.~B.}\ \bibnamefont
  {Abelev}} \emph {et~al.} (\bibinfo {collaboration} {ALICE Collaboration}),\
  }\bibfield  {title} {\bibinfo {title} {{Measurement of quarkonium production
  at forward rapidity in $pp$ collisions at $\sqrt{s} = 7$ TeV}},\ }\href
  {https://doi.org/10.1140/epjc/s10052-014-2974-4} {\bibfield  {journal}
  {\bibinfo  {journal} {Eur. Phys. J. C}\ }\textbf {\bibinfo {volume} {74}},\
  \bibinfo {pages} {2974} (\bibinfo {year} {2014})}\BibitemShut {NoStop}%
\end{thebibliography}
 
%
 
\end{document}